\documentstyle[epsfig]{jaa}

\DeclareMathAlphabet{\mathsc}{OT1}{cmr}{m}{sc}
\def\testbx{bx}%
\DeclareRobustCommand{\ion}[2]{%
\relax\ifmmode
\ifx\testbx\f@series
{\mathbf{#1\,\mathsc{#2}}}\else
{\mathrm{#1\,\mathsc{#2}}}\fi
\else\textup{#1\,{\mdseries\textsc{#2}}}%
\fi}
\def\ch{\footnotesize}
\def\HI{\ion{H}{i}~}
\def\km{km~s$^{-1}$~}
\def\deg{\hbox{$^\circ$}~}
\def\aa{Astron. Astrophys.}
\def\ApJ{Astrophys. J.}
\def\MNRAS{Mon. Not. R. Astron. Soc.}
\def\AJ{Astron. J.}
\def\ARAA{Ann. Rev. Astron. Astrophys.}
\begin{document}
\title[Eridanus group : \HI observations]
{GMRT \HI observations of the Eridanus group of galaxies}
\author[Omar \& Dwarakanath] 
{A. Omar\thanks{Present address : ARIES, Manora peak, 
Nainital, 263 129, Uttaranchal, India}\thanks{e-mail: aomar@upso.ernet.in}
\& K.S. Dwarakanath
\thanks{e-mail: dwaraka@rri.res.in}\\
Raman Research Institute, Sadashivanagar, Bangalore 560 080, India \\}

\pubyear{xxxx}
\volume{xx}
\date{Received xxx; accepted xxx}
\maketitle
\label{firstpage}
\begin{abstract}

The GMRT \HI 21cm-line observations of galaxies in the Eridanus group are presented. The
Eridanus group, at a distance of $\sim23$~Mpc, is a loose group of $\sim200$ galaxies. The
group extends more than 10~Mpc in projection. The velocity dispersion of the galaxies in the group is
$\sim240$~\km. The galaxies are clustered into different sub-groups. The overall
population mix of the group is 30\% (E+S0) and 70\% (Sp+Irr). The observations of 57
Eridanus galaxies were carried out with the GMRT for $\sim200$ hour. \HI emission was detected
from 31 galaxies. The channel rms of $\sim1$ mJy~beam$^{-1}$ was achieved for most of the
image-cubes made with 4 hour of data. The corresponding \HI column density sensitivity
($3\sigma$) is $\sim1\times10^{20}$~cm$^{-2}$ for a velocity-width of $\sim13.4$~\km. The
$3\sigma$ detection limit of  \HI mass is $\sim1.2 \times 10^{7}$~M$_{\odot}$ for a
line-width of 50~\km.  Total \HI images, \HI velocity fields, global \HI line profiles,
\HI mass surface densities, \HI disk parameters and \HI rotation curves are presented.
The velocity fields are analysed separately for the approaching and the receding sides of
the galaxies. This data will be used to study the \HI and the radio continuum properties,
the Tully-Fisher relations, the dark matter halos, and the kinematical and \HI
lopsidedness in galaxies. 

\end{abstract} 


\section{Motivation}

Several redshift surveys carried out over the past several decades indicate that
galaxies are distributed inhomogeneously in the local Universe. The regions of
highest galaxy densities are superclusters and clusters. However, the majority
of galaxies in the local Universe are found in less dense regions called
groups. According to theories of hierarchical structure formation, clusters are
built via mergers of groups.  Clusters differ from groups in several aspects. A
remarkable difference is observed in the morphological mix and \HI content of
the galaxies. Clusters have an enhanced population of both the early type (S0
and E) galaxies and the \HI depleted spirals (Curtis 1918, Hubble \& Humason
1931, Davies \& Lewis 1973, Giovanelli \& Haynes 1985, Warmels 1988, Cayatte et al. 1990,
Bravo-Alfaro et al. 2000) while groups are populated mainly by \HI rich
spirals.  Dressler (1980) noticed a tight correlation between the galaxy
morphology and the local projected galaxy density. This correlation, known as
the density-morphology relation, is observed to be valid over more than five
orders of magnitude in the projected galaxy density (Postman \& Geller 1984).
The origin of the enhanced population of E+S0's in high galaxy density regions
has been the subject of much debate. There are two hypotheses for the formation
of S0's, one is ``Nature'' where it is believed that early types were formed as
such, and the other is ``Nurture'' according to which S0's are transformed spirals.
Some of the recent observations indicate that the clusters at intermediate
redshifts ($z\sim0.1 - 0.3$) tend to have a higher fraction of S0's at the
expense of spirals (e.g., Poggianti et al. 1999, Dressler et al. 1997, Fasano
et al. 2000). These observations support the ``Nurture'' scenario.

Several gas-removal mechanisms, viz., ram-pressure stripping (Gunn \& Gott 1972), thermal
conduction (Cowie \& Songaila 1977), viscous stripping (Nulsen 1982), harassment (Moore
et al. 1998), starvation etc. have been proposed to explain the \HI deficiency in
cluster spirals. Some of these processes are also believed to be driving the transformation
of spirals to S0's. Each of these processes has been predicted to remove \HI mass of the
order of the typical \HI mass of a galaxy. These processes, however, are quite sensitive to
several parameters like the density of the intra-cluster medium (ICM), the radial velocity
vector of the galaxy, the magnetic field in the ICM, the gas-reservoir in the galaxy halo, etc..
Some of these parameters have poor estimations which make the efficacy of these processes
doubtful. It has been argued that no single gas-removal mechanism can explain the global
\HI deficiency in cluster spirals (e.g, Magri et al. 1988). The exact physical
mechanism(s) responsible for \HI depletion in cluster spirals, therefore, remains
uncertain. These difficulties have led one to speculate that cluster galaxies were
perhaps \HI deficient even before they fell into the cluster. Such a scenario can be
explored by studying groups of galaxies.

Several groups have been previously imaged in \HI, e.g., the Hickson Compact Groups (HCGs;
Verdes-Montenegro et al. 2001) and the the Ursa-Major group (Verheijen \& Sancisi 2001).
The galaxy densities in HCGs are comparable to that in galaxy clusters, although HCGs
have far less number of galaxies compared to that in clusters.  The galaxies in some of
the HCGs were found to be significantly \HI depleted. HCGs also tend to have a
significant population of early type galaxies. The Ursa-Major group, which has only a
few S0's and no ellipticals, showed no significant \HI deficiency. The environment in the
Ursa-Major group is similar to that in field. Here, we present an \HI survey of the
Eridanus group of galaxies with the recently completed Giant Meterwave Radio Telescope
(GMRT). The  Eridanus group is believed to be at an evolutionary stage intermediate to
that of field and a cluster. The Eridanus group has a significant population of early
type galaxies. The sub-grouping of galaxies in the group is quite prominent. The Eridanus
group also has weak diffuse x-ray emission centered around some of the brightest galaxies in
the sub-groups. On a broader perspective, the properties of the Eridanus group are
between that of a loose group like the Ursa-Major and a cluster like Fornax or Virgo. 
The main aim of
this survey is to identify the galaxy evolution processes active in an environment
intermediate between that of a cluster and field.

The GMRT observations provided both the \HI and the radio continuum ($\nu \sim 1.4$~GHz)
data. The kinematical information of galaxies has also been obtained using the \HI data.
This survey has capabilities to carry out several other studies. Some of the studies
proposed to be carried out are the following - 

\begin{itemize}
\item \HI content of galaxies in the Eridanus group.

\item \HI morphologies of galaxies in the group.

\item Tully-Fisher relations.

\item Radio -- Far-infrared correlation.

\item Rotation curves and dark matter halos.

\item Kinematical and \HI lopsidedness.

\end{itemize}

In the present paper, the GMRT observations and the data analyses are described. We also
investigate correlations between \HI and optical properties in this paper. The paper is
arranged in the following order. The next section describes the properties of the
Eridanus group. Sect.~3 contains details of the GMRT observations. The analyses of the
\HI images are described in Sect.~4. Some of the \HI  properties of the Eridanus galaxies
are discussed in Sect.~5. The results are presented in the tables in App.~A. The \HI
atlas consists of the \HI images, the \HI velocity fields, the global \HI profiles, the
\HI surface densities, the \HI rotation curves, and the kinematical parameters of the \HI
disks. The \HI atlases are given in App.~B.

\section{The Eridanus group}

\subsection{Introduction}

The concentration of galaxies in the Eridanus region is known for many decades (Baker
1933, 1936). The complex morphology of this region was pointed out by  de Vaucouleurs
(1975). The Eridanus group was identified as a moderate size cluster in a large scale
filamentary structure near $cz\sim1500$~\km in the Southern Sky Redshift Survey (SSRS; da
Costa et al. 1988). This filamentary structure, which is the  most prominent in the
southern sky, extends for more than 20~Mpc in projection. The Fornax cluster and the Dorado group of
galaxies are also part of this filamentary structure. The dynamical parameters of a few
galaxies in the Eridanus group were first published by Rood et al. (1970). With the
increased number of identifications in this region by Sandage \& Tammann (1975) and Welch
et al. (1975), the latter authors speculated a dynamical connection between the Fornax
cluster and the Eridanus group.  Using the data from the Southern Galactic Cap sample
(SGC; Pellegrini et al. 1990), Willmer et al. (1989)  grouped the galaxies in the
Eridanus region into different sub-groups and studied their dynamics. They concluded that
each sub-group is a bound structure and possibly the entire group is also gravitationally
bound with a dynamical mass greater than $10^{13}$~M$_{\odot}$. They further pointed out
that the Fornax and the Eridanus together constitute a bound system.  The Eridanus group is a
dynamically young system with a velocity dispersion of $\sim240$~\km, which is lower
compared to that ($\sim1000$~\km) seen in clusters like the Coma. The distance to the group
is estimated as $23\pm2$~Mpc based on the surface brightness fluctuation measurements
(Tonry et al. 1997, Jensen et al. 1998, Tonry et al. 2001). All identified members in the
group are in the Helio-centric velocity range of $\sim1000 - 2200$~\km, except NGC~1400 (S0),
which has a velocity of $\sim558$ \km. However, NGC~1400 is predicted to be at a similar
distance as that of the other members of the group based on the surface brightness fluctuation
measurements.

\subsection{Group structure, membership, and morphological mix}

\begin{figure}
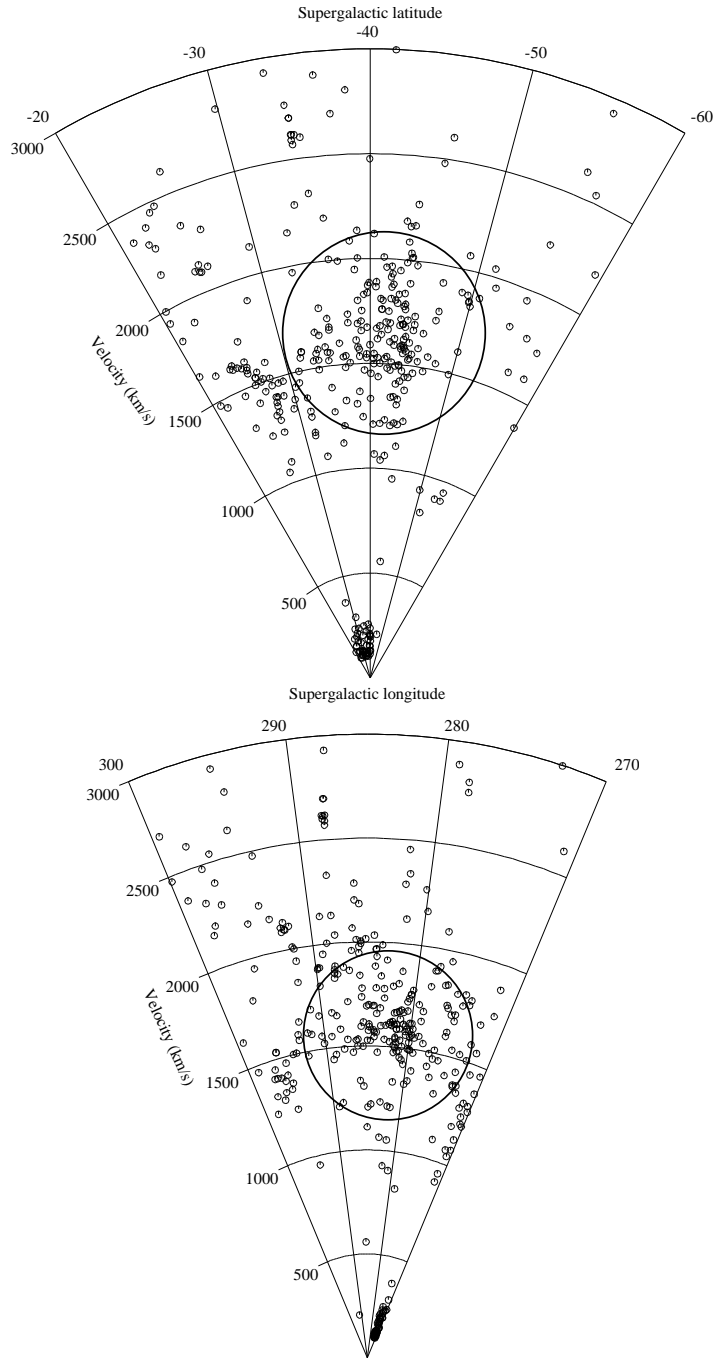
 
\centering 
\includegraphics[width=9cm, angle=-90]{cone2.2.epsi}
\includegraphics[width=9cm, angle=-90]{cone1.2.epsi} 

\caption{Positions of galaxies in the velocity-cone diagrams. The velocities
are Heliocentric from the NASA Extra-galactic Database (NED). The circles mark
the Eridanus group. } 

\label{fig:cone} 
\end{figure}

\begin{figure}
\centering
\includegraphics[width=12cm, angle=0]{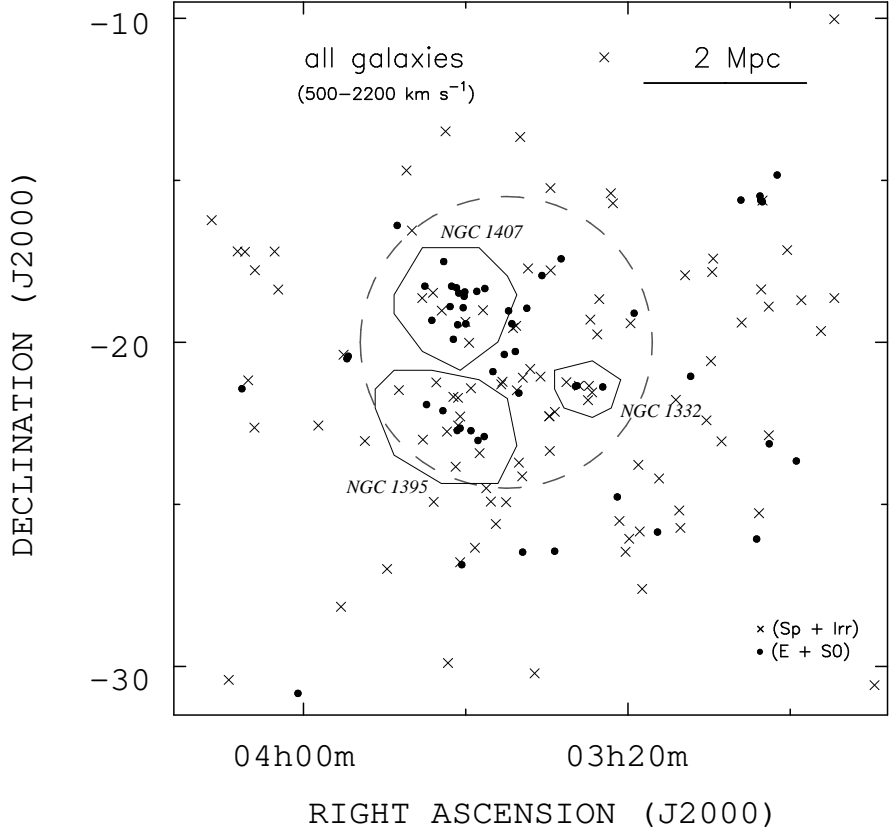}

\caption{Galaxies in the Eridanus group. The radius of the dashed circle is
$\sim2$~Mpc within which galaxies were observed with the GMRT. A few sub-groups
are marked with their approximate boundaries as identified by Willmer et al.
(1989).}

\label{fig:l-b}
\end{figure}
\begin{figure}
\centering
\includegraphics[width=7.5cm, angle=-90]{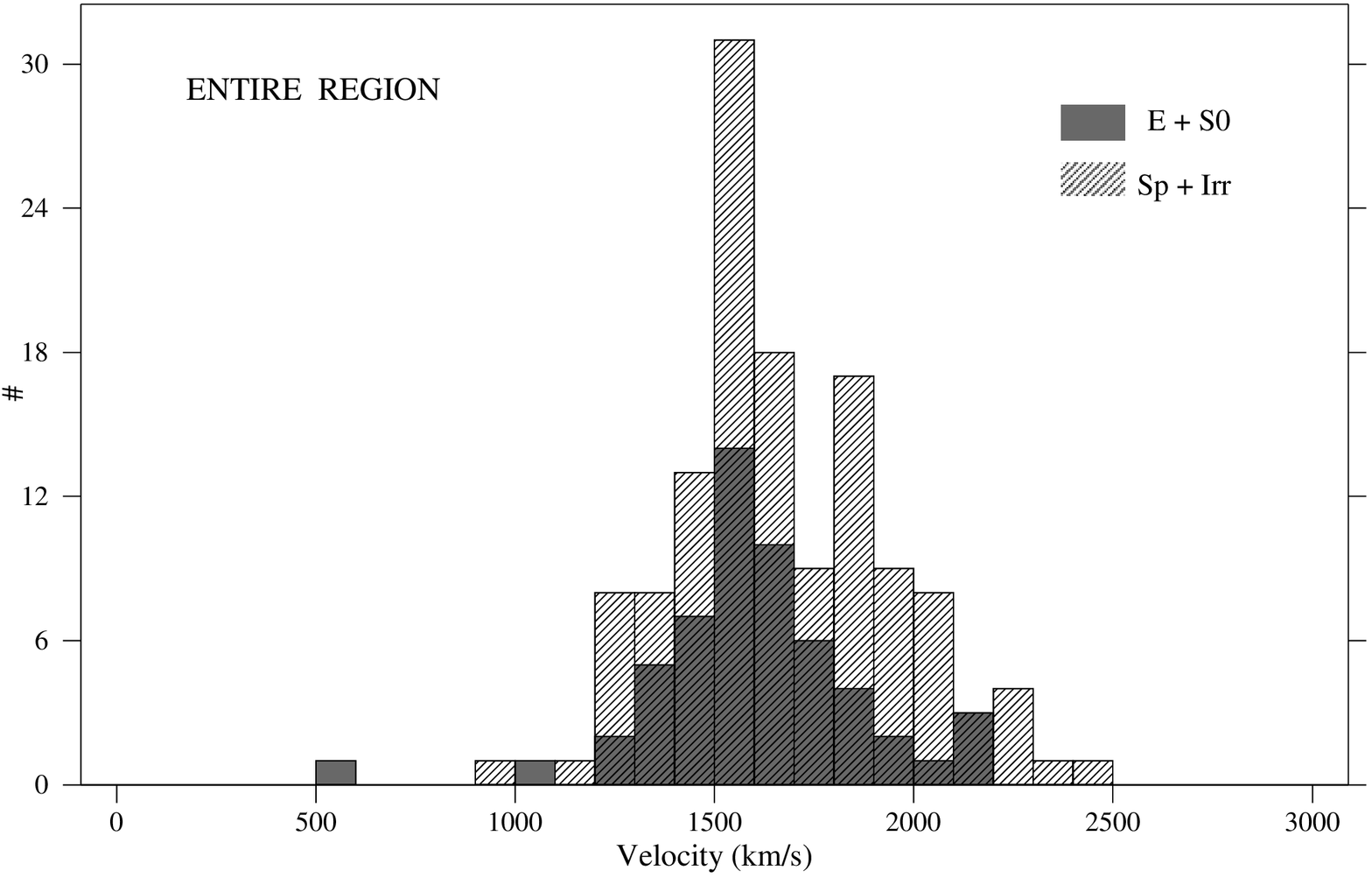}
\includegraphics[width=7.5cm, angle=-90]{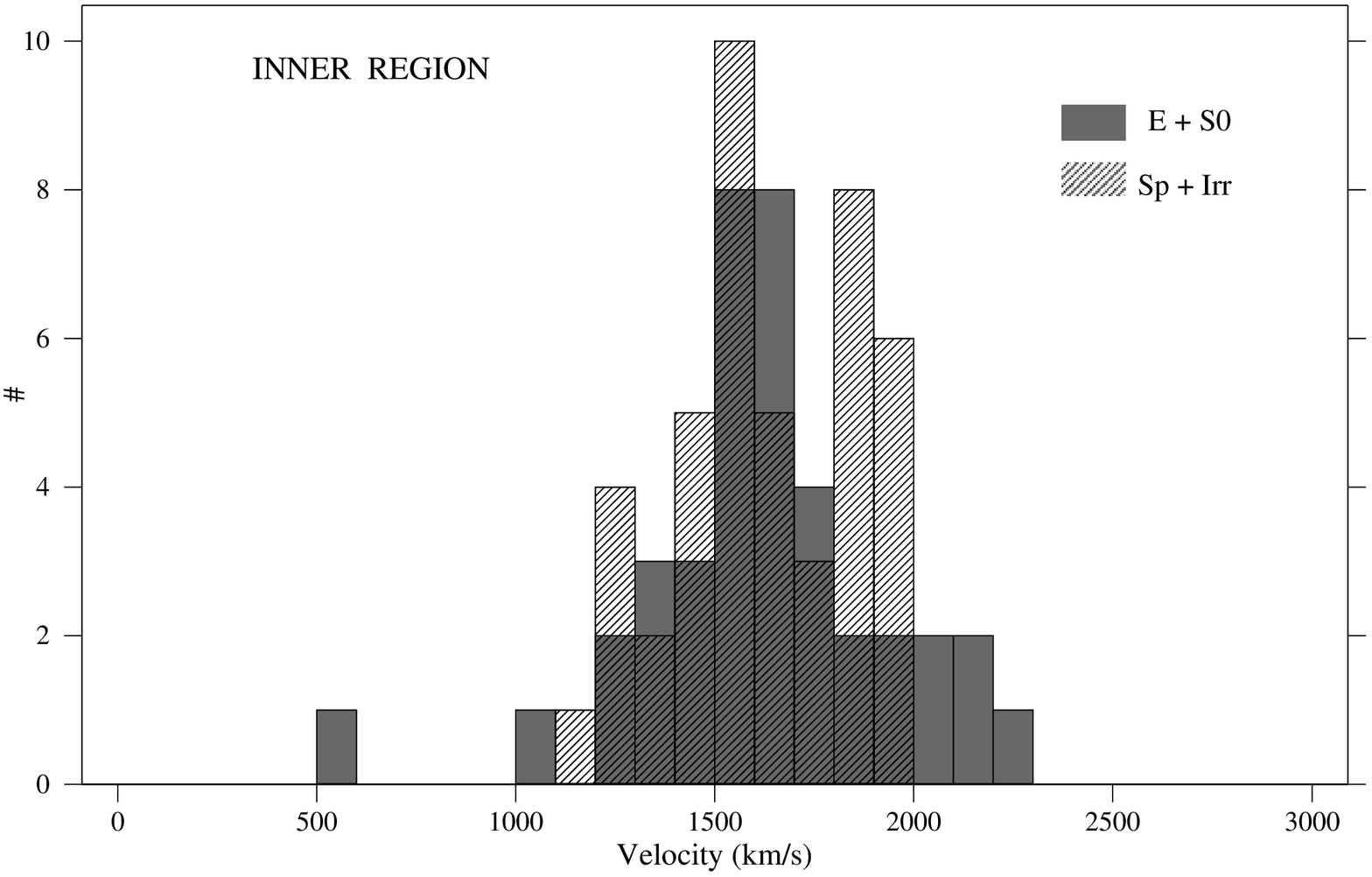}

\caption{Velocity-histograms of galaxies in  the inner 4~Mpc region (bottom
panel) and in the entire group (top panel) are shown. There is no appreciable
difference in the velocity range over which the early types and the late
types are distributed. }

\label{fig:histo}
\end{figure}

The velocity-cone diagrams are plotted in Fig.~\ref{fig:cone}. The plots are in the
Supergalactic coordinates. The velocities obtained from the NASA Extra-galactic Database
(NED) are Heliocentric, and follow the optical definition. The clustering of
galaxies near  $l=283\deg$ and $b=-43\deg$ is the Eridanus group.  Most of the galaxies
are concentrated in the velocity range $cz = 1000-2200$~\km. The group appears to be
loose and irregular. The clustering of galaxies near the apex ($cz = 0$) is the local
group. In Fig.~\ref{fig:l-b}, the positions of galaxies within the velocity range
$500-2500$~\km are plotted.  There are 181 galaxies in this plot, 60 early types (E+S0)
and 121 late types (Sp+Irr). The approximate boundaries of three main sub-groups
identified by Willmer et al. (1989) are marked in this figure. The sub-clustering of
galaxies can be seen in this figure. 

It can be seen that most of the early type galaxies are in the region inside the circle marked in
Fig.~\ref{fig:l-b}. The sub-clustering is also prominent in the inner region. In the outer regions,
population is dominated by spirals.  The morphological mix  is appreciably different in each sub-group.
The three sub-groups namely NGC~1407, NGC~1332, and NGC~1395 have their brightest members as an
elliptical or an S0. The NGC~1407 sub-group is the richest in the early types, most of them being S0s.
The population of (E+S0) and (Sp+Irr) in the NGC~1407 sub-group is 70\% \& 30\% respectively, while that
in most of the other sub-groups is $\sim$40\% \& 60\% respectively. The overall population mix of the
Eridanus group is $\sim$30\% (E+S0) \& 70\% (Sp+Irr). 
The velocity-histograms of the early types and the
late types are plotted in Fig.~\ref{fig:histo}. 
There is no appreciable difference in the velocity range
over which the early types and the late types are distributed. 
However, it can be seen from the upper panel of Fig.~3 that the
distribution of early-type galaxies is approximately a Gaussian whereas the late-type galaxies are
rather uniformly distributed. Upon inspection of the locations of late-type galaxies at lower (1000-1300
\km) and higher (1800-2100 \km) velocity ends from the mean, it appears that the galaxies having higher
velocities are more uniformly distributed in the sky compared to those at lower velocities. The lower
velocity galaxies are largely confined within the circle drawn in Fig.~2. Further, it is interesting to
note that the population mix of the NGC~1407 sub-group is similar to that seen in evolved clusters like
the Coma, whereas the velocity dispersion ($\sim$ 250 \km) of the NGC~1407 sub-group is much smaller than
that of Coma ($\sim 1000$~\km). Some other groups like Leo I, NGC 3607, and NGC 5846, which have lower
velocity dispersions compared to that in clusters are also populated mainly by early type galaxies.

\subsection{X-ray properties}

\begin{figure}
\centering
\includegraphics[width=9cm, angle=0]{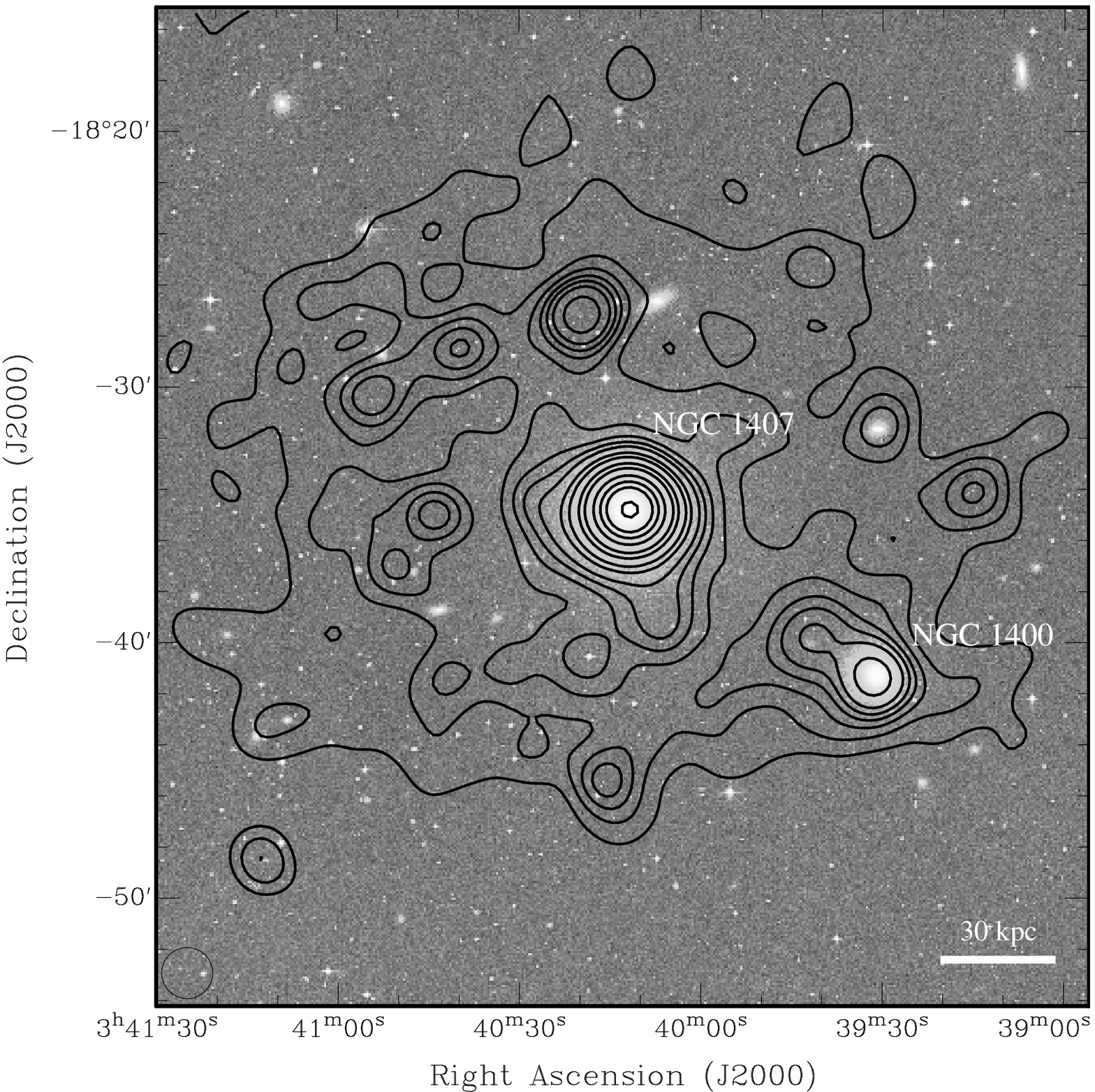}
\includegraphics[width=9cm, angle=0]{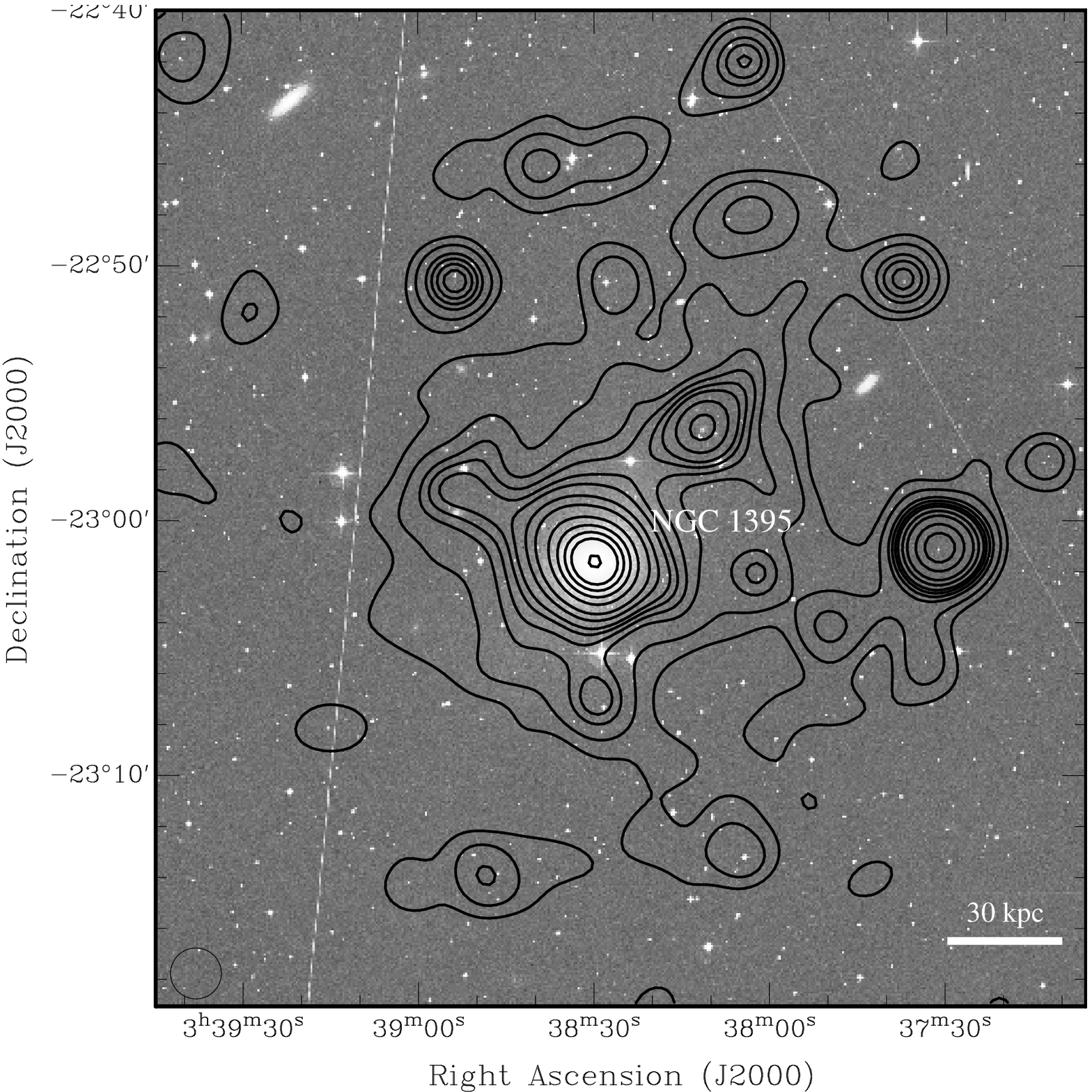}

\caption{Contours of X-ray emission around NGC~1407 and NGC~1395 overlaid upon the
 optical images from the DSS. The X-ray images are retrieved from the ROSAT PSPC
 archived data and smoothed with a  circular Gaussian beam of $90"$.}

\label{fig:xray}
\end{figure}

The optically bright early type galaxies NGC~1400, NGC~1407, NGC~1395, and  NGC~1332 are
known X-ray sources in this group.  Trinchieri et al. (2000) reported the presence of
diffuse X-ray emission around NGC~1407. The processed and calibrated  X-ray images (0.1
keV - 2.0 keV) centered at NGC~1407, NGC~1395, and NGC~1332 were obtained from the ROSAT
PSPC (Roentgen Satellite Position Sensitive Proportional Counter) archival data. Each
field was observed for $\sim6$ hour using the ROSAT PSPC instruments. The soft X-ray
images shown in Fig.~\ref{fig:xray} were convolved with a circular Gaussian  beam of FWHM
$90''$ to enhance the diffuse emission.  Apart from the X-ray emission associated with
NGC~1400, NGC~1407 and a few other unresolved sources in the field, diffuse emission
centered at NGC~1407 and NGC~1395 can be seen in Fig.~\ref{fig:xray}.  The extent of the
diffuse emission is $\sim30'$ ($\sim200$~kpc) around NGC~1407 and $\sim20'$
($\sim135$~Kpc) around NGC~1395. No diffuse emission is seen around  NGC~1332 (not shown
here). Using the PIMMS (Portable, Interactive Multi-Mission Simulator; Mukai 1993) tool,
the diffuse emission was modeled as thermal free-free emission from a Raymond-Smith
plasma of energy $\sim1.0$~keV ($T\sim10^{7}$~K) and metallicity 0.2--solar. The choice
of the temperature and the metallicity is in accordance with typical values found in
X-ray groups (Mulchaey 2000). 

The total X-ray luminosity of the diffuse emission in the energy range $0.1 - 2.0$~keV is
$1.6 \times 10^{41}$ erg~s$^{-1}$ for the NGC~1407 sub-group and  $\sim6.8 \times
10^{40}$ erg~s$^{-1}$ for the NGC~1395 sub-group. The intra-group medium density is
estimated as $\sim2.0 \times 10^{-4}$ cm$^{-3}$ in the X-ray emitting region.
The X-ray luminosity of the Eridanus group is about 2-3 orders of
magnitude lower compared to that of the clusters like Coma and Virgo.
The estimated intra-group medium
density is about an order of magnitude lower than that observed in virialised clusters
like Coma. 

\subsection{Comparison of the Eridanus group with other groups and clusters}

The properties of the Eridanus group are compared with the Virgo and the Fornax cluster,
and the Ursa-Major group in Tab.~\ref{tab:compare}. All these systems are at comparable
distances. Both the Fornax cluster and the Eridanus
group belong to the filamentary structure described by da Costa et al. (1988).  The
Ursa-Major group is a loose group of galaxies (Tully et al. 1996). All the four systems
have quite different properties. The Fornax cluster having the highest galaxy density has
the lowest spiral fraction, consistent with the density-morphology relation. 
The Eridanus group is intermediate between
the Virgo cluster and the Ursa-Major group in terms of its velocity dispersion, 
its x-ray luminosity, and its number of early-type galaxies.  The
mean projected galaxy density in Eridanus is intermediate between that in Ursa-Major and
in Virgo. The galaxies in Virgo are \HI deficient. The Ursa-Major group has normal \HI
content. The X-ray luminosity of the Eridanus group is at the lower end of the X-ray
luminosities observed in groups (Mulchaey 2000). The velocity dispersion of the galaxies
in the Eridanus group is intermediate between that in Fornax and in Ursa-Major. From this
comparison, it appears that the Eridanus group forms a system which is intermediate
between a loose group (Ursa-Major) and a rich cluster (Virgo and Fornax).

\begin{table}
\begin{center}
\caption{Comparison of four nearby galaxy groups and clusters}
\vspace{0.1in}
\label{tab:compare}
\begin{tabular}{lcccc}
\hline
\hline
\bf {Properties} & \bf{Virgo$^{a}$} & \bf{Fornax$^{b}$} & \bf{Eridanus$^{c}$} & \bf{Ursa-Major$^{d}$} \\
\hline
 & & & & \\
Distance (Mpc)			  & 17   & 20     & 23      & 21  \\
No. of E+S0's                     & 71   & 23     & 36      & 9   \\
No. of Sp + Irr's                 & 123  & 17     & 42      & 53  \\
 (S+Irr) fraction		  & 0.6  & 0.4    & 0.5     & 0.9 \\
Vel. dispersion (~\km)            & 760  & 350    & 240     & 150 \\
log X-ray luminosity (erg s$^{-1}$)& 43.5 & 41.7   & 41.4    & -- \\
References			  &{\ch 1,2,3,4} &{\ch 2,5,6,7}  &{\ch 8}  &{\ch 9,10} \\
& & & & \\
\hline 

\multicolumn{5}{p{5.5in}}{\ch  Notes - (a): Inner 6\deg region, (b): Inner 2\hbox{$^\circ$}.4
region, (c): Inner 9\deg region, (d): Inner 15\deg region. References - (1) Federspiel et
al.(1998), (2) Ferguson (1989), (3) Binggeli et al. (1987), (4) Mushotzky \& Smith (1980),
(5)Mould et al. (2000), (6) Richter \& Sadler (1985), (7) Paolillo et al. (2001), (8) Omar
(2004), (9) Sakai et al. (2000), (10) Tully et al. (1996) }

\end{tabular} 
\end{center} 
\end{table}

\section{Observations and data reduction}

The present GMRT \HI observations offer several advantages over studies carried out in
the past  using single dish telescopes.  The GMRT is an interferometric array of thirty
45-m diameter fully steerable parabolic dishes. A description of the GMRT is given by
Swarup et al. (1990). The GMRT is located at a site (longitude = $74\deg.05$~E, latitude
= $19\deg.092$~N, height $\sim650$m above MSL)  about 80~km north of Pune, India. The
configuration of the GMRT is optimized to meet the requirements of high angular
resolution and of being able to image extended emission. This
optimization is achieved through a hybrid configuration of the antennas. Fourteen of the
thirty dishes are located more or less randomly in a compact central array within an area
of about  $1\times1$~square kilometer, and the remaining sixteen dishes are spread out
along the 3 arms of an approximately ${\sf Y}$ shaped configuration over a larger region.
The longest separation of antennas is $\sim25$~km, and the shortest separation is
$\sim100$~m. The GMRT is expected to be sensitive to structures on the scales of $2'' - 7'$
at a wavelength of 21~cm. The angular sizes of the galaxies in the Eridanus group are
in the range $1' - 5'$ implying that the data should be sensitive to image radio emission
(\HI and continuum) over the full extents of galaxies. The FWHM of the primary beam of a
GMRT antenna is $\sim24'$ at 1.4~GHz.

\subsection{Sample of galaxies}

\begin{figure}
\centering
\includegraphics[width=12cm, angle=0]{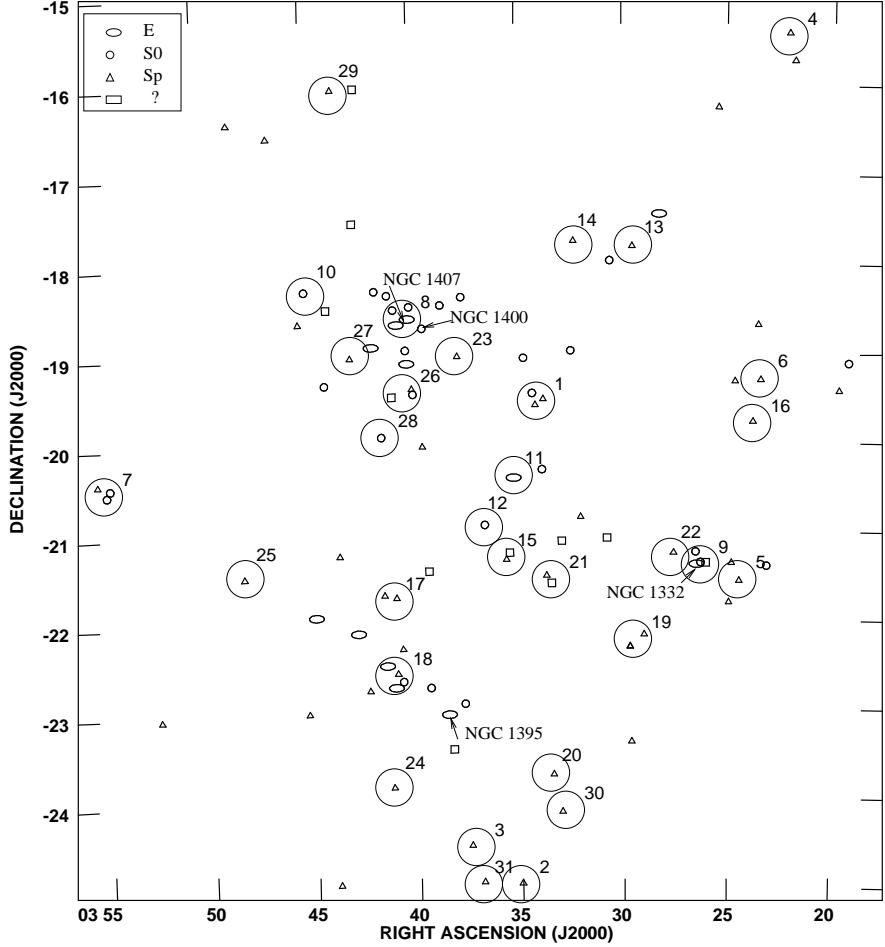}

\caption{The fields observed with the GMRT. The bigger circles correspond to the FWHM of
the GMRT primary beam ($\sim24'$) at 1.4~GHz.}

\label{fig:obs}
\end{figure} 

The selection of galaxies for the \HI observations were made keeping in mind the broad
perspective of the work. The galaxies were not selected based on their \HI contents or
their optical luminosities. Both early type and late type galaxies were included in the
sample.  The galaxies were selected from the inner 4~Mpc region of the group where galaxy
density is higher and most of the S0's are found. A follow up R-band photometric
observations were also carried out with the 1-m optical reflector at the Aryabhatta
Research Institute of Observational Sciences (ARIES; $formerly$ State Observatory),
Nainital. The optical data analysis is presented in Omar (2004). 

Since the present study was carried out with a limited telescope time of $\sim200$ hour,
the pointing centres of the observations were adjusted in a way to include two or more
galaxies within the FWHM of the primary beam. Unfortunately, one complete run of
observations on 16 galaxies (during November 2001), mostly early types, 
was badly affected due
to ionospheric scintillations, perhaps related to the intense solar activities during that
year. The data collected during this period could not be used to obtain images. Five
galaxies from these lost observations were re-observed later in 2002. The science quality
data were obtained for a total of 46 galaxies. In Tabs~\ref{tab:sample} \&
\ref{tab:prop}, the complete observed sample of 57 galaxies is listed with some of their
previously known optical and radio properties.

\subsection{Observational parameters}

The Eridanus group can be observed with the GMRT for $\sim8$ hour in a given day. Often,
two galaxies were observed in each day. The observing strategy  was optimized to get
uniform distribution of  visibilities. Two galaxies were observed alternately for
$15-20$ minute each followed by $5-7$ minute of observations of secondary calibrators.
This cycle was repeated and a total of 3-4 hour of observing time was accumulated on each
galaxy. Most of the observations were carried out using an 8~MHz bandwidth over 128
channels, which gives a velocity resolution of $\sim13.4$~\km. The observations were
carried out for longer duration ($\sim8$ hour) for some of the early type galaxies, and
with smaller bandwidths (2-4~MHz) for smaller inclination galaxies to get sufficient
velocity resolution. A total of $\sim200$ hour of the GMRT observations were carried out
spread over a period of 2 years (2000 -- 2002). The data obtained during November, 2001
which were corrupted due to scintillations were discarded. The observing parameters are
listed in tab.~\ref{tab:Radobs}.

The VLA calibrators 0240-231 and 0409-179 were used as the secondary calibrators.
0240-231 is classified as ``un-resolved'' for all the four VLA configurations with a
20-cm flux density of 6.3~Jy. 0409-179 is resolved by baselines longer than 10~km with a
20-cm flux density of 2.2~Jy. 0137+331 (3C~48) and 0542+498 (3C~147) were used as the
primary calibrators. 3C~48 is resolved by baselines longer than 8~km with a 20-cm flux
density of 16.5~Jy and 3C~147 is resolved by baselines longer than 10~km with a 20-cm
flux density of 22.5~Jy. 3C~48 was observed in the beginning and 3C~147 was observed at
the end of each observing run for 20-30 minutes. The flux densities of the primary
calibrators were estimated at the observed frequencies using their known radio spectra
from the VLA observations in the 1999.2 epoch.

\subsection{Data acquisition and reduction}

The data (visibilities) were collected in the {\sf LTA} ({\it Long Time Accumulation})
format, which is the native format for the GMRT data. The {\sf LTA} data were converted
to {\sf FITS} ({\it Flexible Image Transport System}) format for subsequent processing.
Visibilities were averaged over $\sim16$~second. The data were monitored on line. The data
were later flagged from the antennas having low gains, for time ranges where data were
corrupt, and at lower  elevations (usually below $25\deg$) where correlation drops
significantly (below 50\% in some cases). The flux densities of the secondary calibrators
were estimated based on the flux densities of the primary calibrators. The
visibilities on the target galaxies were calibrated by interpolating  the complex gains
determined using the secondary calibrators. Since the spectral responses of filters are
not flat, the initial calibration was carried out using the data averaged over four to six
channels. The  spectral response of the antennas were determined using both the secondary
and the primary calibrators, and an averaged spectrum was used to correct the band
shapes. The gains start declining significantly after the 110$^{th}$ channel. The first
$1-3$ channels are generally corrupted in the filter response. Therefore, the data were
used between channels $3-115$.  An initial \HI spectrum was generated using the {\sf
AIPS} ({\it Astronomical Image Processing System}) task {\sf POSSM} at spatial
frequencies below 2~k$\lambda$ in the direction of target galaxies. This range of spatial
frequencies enables most of the \HI signal to be detected in the \HI spectrum. This
spectrum is to identify channels with \HI emission. 

The continuum-data were generated by averaging the channels devoid of \HI line emission.
The continuum images were made using this channel averaged data and were used for
self-calibration.  Several iterations of phase self-calibrations were performed to
improve the dynamic range of the images. The final self-calibrated solutions were applied
to the line-data. The continuum emission was subtracted from the line-data using the {\sf
AIPS} tasks {\sf UVSUB} and {\sf UVLIN}. The self-calibrated and continuum subtracted
line-data were used to make the image cubes at different resolutions by selecting
appropriate $(u,v)$ ranges. The image cubes were made at two resolutions - one 
with a resolution of $\sim15''$
(high resolution cube) using $(u,v)$ data in the range $0.2 - 20 k\lambda$, and another with
a resolution of 
$\sim50$'' (low resolution cube) using $(u,v)$ data in the range $0.2 - 5 k\lambda$.

\section{Image analysis}

The images were analysed using the {\sf GIPSY} ({\it Groningen Image Processing System})
package developed by the Kapteyn Institute, the {\sf KARMA} visualization tool (Gooch 1996), 
and the {\sf
AIPS} package developed by the National Radio Astronomy Observatory.

\begin{figure}
\centering
\includegraphics[width=14cm, angle=0]{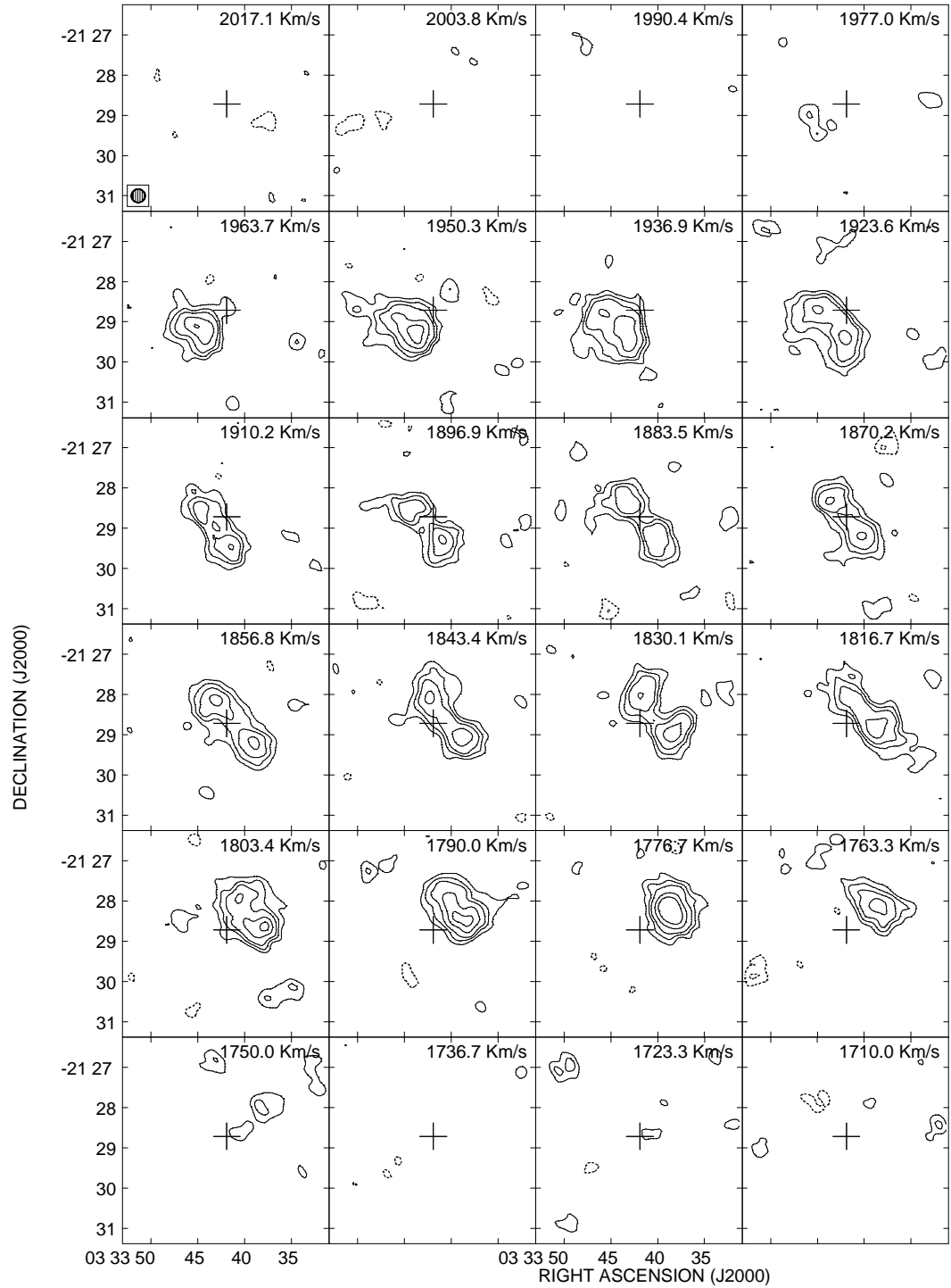}

\caption{\HI emission from IC~1953. The `+' sign marks the optical centre of the galaxy.
The rms/channel is 1.4 mJy~beam$^{-1}$. The contours are at 2.5, 3.75, 5, 7.5, and 10
times the rms.  The images
are convolved with a circular Gaussian beam of $20'' \times 20''$.}

\label{fig:chan}
\end{figure}

Since the angular resolution varied by a few arc second in different cubes, all high
resolution cubes were convolved to a common resolution of $20'' \times 20''$. In some
cases, intermediate resolution cubes at $25''$ or $30''$ were also made. The channel
images typically have an rms of $1$ mJy beam$^{-1}$.   The $3\sigma$ column density
detection limit in the $20''$ images is $1 \times10^{20}$ cm$^{-2}$. The cubes are
sensitive (3$\sigma$) to detect a galaxy of \HI mass $1.2\times10^{7}$ M$_{\odot}$ for an
\HI line-width of 50 km s$^{-1}$.  The image cubes were inspected visually to identify
\HI signals.  The channel images are presented for all the \HI detected galaxies
elsewhere (Omar 2004). An example of the channel images is shown in Fig.~\ref{fig:chan}.

\begin{figure}
\centering
\includegraphics[width=12cm, angle=0]{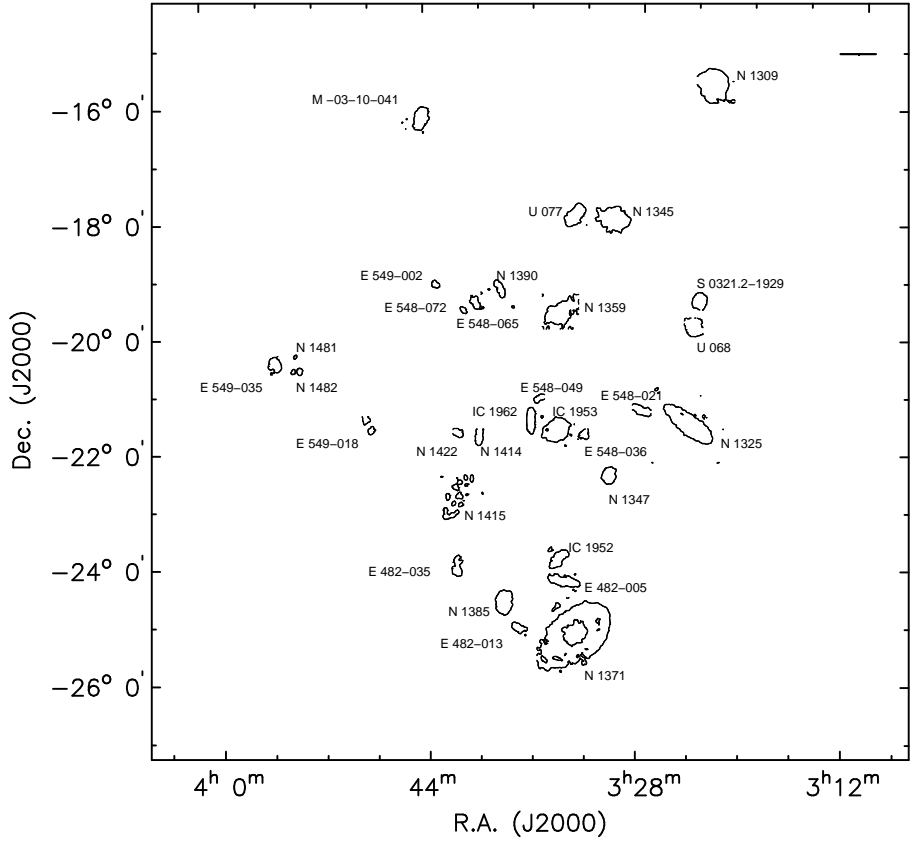}

\caption{A contour-image collage of the \HI detected galaxies in the Eridanus
group. Only one contour is plotted to indicate the extent of each galaxy at
$N_{\HI} = 10^{20}$~cm$^{-2}$. The individual galaxies are magnified ten
times. To avoid overlap, some galaxies are slightly displaced from their actual
positions. A bar at the upper right hand corner indicates a scale of 20~kpc for
the enlarged sizes of the galaxies. Otherwise, $1\deg$ corresponds to
$\sim400$~kpc. }

\label{fig:contview}
\end{figure}

\begin{figure}
\centering
\includegraphics[width=12cm, angle=0]{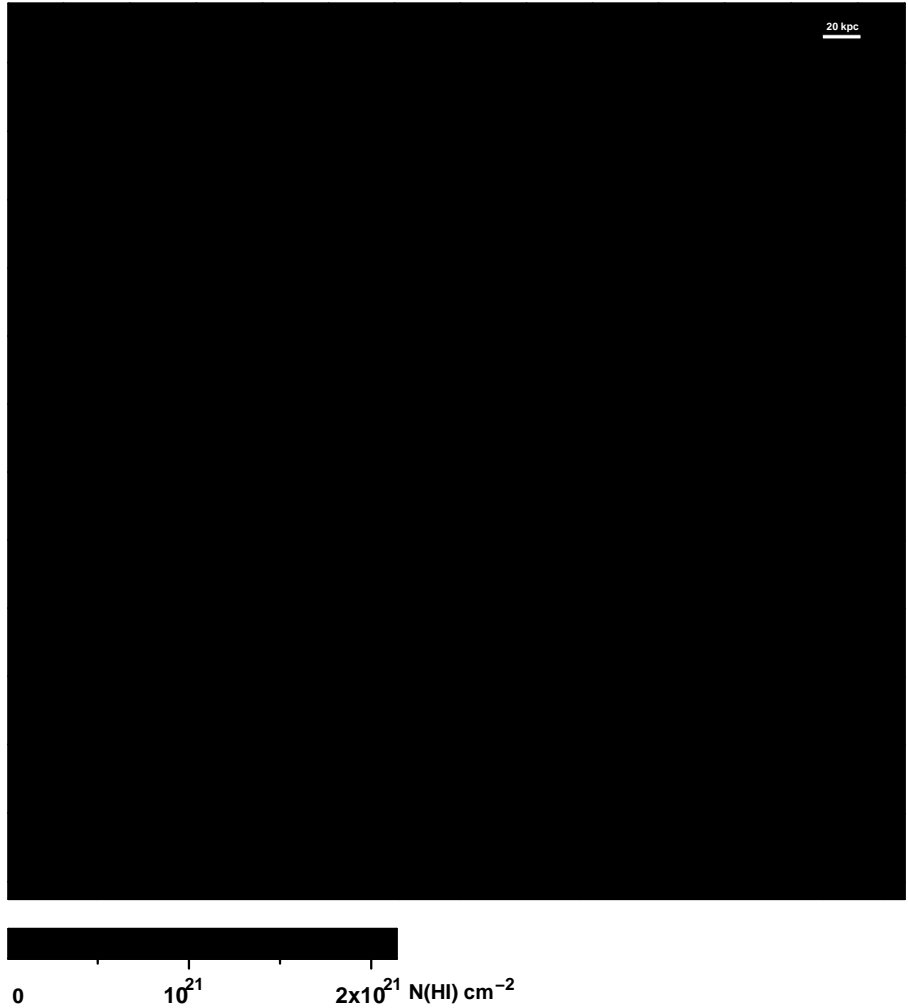}

\caption{A collage of the \HI detected galaxies in the Eridanus
group. The color key indicates the \HI column density. }

\label{fig:grpview}
\end{figure}

\subsection{Total \HI map and \HI diameter}

The zeroth and the first order moment maps were generated at both  the low ($50''$) and 
the high ($20''$) angular resolutions. The moment zero map or the total \HI image is
obtained by summing the \HI images in different channels. The cubes were first blanked to
separate the \HI signals from noise before summing the channels. The blanking can be done
in several ways. The total \HI image depends on the blanking procedure (Rupen 1999). One
of the methods is to blank the pixels below a certain flux density level. A higher cutoff (e.g.,
5$\sigma$) makes the total \HI images patchy while a lower cutoff (e.g., 3$\sigma$) makes
the images noisy. The low surface brightness nature of \HI emission  makes it difficult to
separate the low level signals from noise. A hybrid approach has been shown to be effective
in overcoming this problem (Rupen 1999). This approach involves masking the noise
after smoothing the cube using the {\sf AIPS} task {\sf MOMNT}. The moment maps are still
estimated using the un-smoothed cube in this approach. 

The flux density (mJy beam$^{-1}$) is converted to \HI column density using the
following relation (eqn. 3.38, Spitzer 1978):

\begin{equation}
N_{\HI}(\alpha,\delta) = \frac{1.1 \times 10^{21} cm^{-2}}{\theta_{a} \times \theta_{b}} ~{\delta}v \sum_{j=1}^{N_{chan}} S_{j}(\alpha,\delta)  
\end{equation}

\noindent Where $\theta_{a}$ and $\theta_{b}$ are the FWHM of the synthesised beam
measured in arc second along the major and minor axes respectively. $S_{j}$ is the \HI
flux density (mJy beam$^{-1}$) in the channel $j$ and ${\delta}v$ is the velocity
resolution in \km. The \HI gas is assumed to be optically thin. A collage of the
integrated \HI maps of the Eridanus galaxies is shown in Fig~\ref{fig:contview} as
contours and in  Fig~\ref{fig:grpview} as colour-coded. The individual galaxies are
enlarged ten times. To avoid overlap, some galaxies are slightly displaced from
their actual positions.

The diameters of the \HI disks were estimated from the high resolution total \HI images
at a fixed face-on \HI surface density of $1$~M$_{\odot}$~pc$^{-2}$. Due to projection
effects, the sensitivities to the face-on \HI surface densities were not uniform.
Therefore, in some cases the \HI diameters were extrapolated to the face-on \HI surface
density level of $1$~M$_{\odot}$~pc$^{-2}$. 

\subsection{\HI velocity field}

The conventional way of deriving the velocity field is to compute the intensity-weighted
first order moment of the \HI images at different velocities. There is an alternative to
obtain velocity fields by fitting the \HI profiles at every pixel with a Gaussian. These
profiles are usually asymmetric depending upon the kinematics of \HI  along the
line-of-sight, and also due to the  beam smearing caused by the finite size of the
synthesized beam. The broadening and the asymmetry will depend upon the \HI distribution
in the galaxy. The effect of beam smearing will be more pronounced in edge-on systems. As
a result of the asymmetry and the broadening in the \HI profile, a single Gaussian
component will not give an accurate result. Unfortunately, multi-component Gaussian fit
could not be carried out as the signal to noise ratio of the detections were not
sufficient.

Rupen (1999) has briefly discussed the merits and drawbacks of these two procedures. In
the present analysis, the first order moment maps were found to be generally noisier than
the velocity field maps obtained by Gaussian fits. This may be because the Gaussian fits
were not as sensitive to the outliers as  moment map were. Therefore, in the present
analysis, Gaussian fits were used to construct the velocity field maps. It should be
noted that both the procedures to obtain the velocity field will underestimate  rotation
velocities at locations of steep velocity gradients. The flat part of the rotation curve,
however, remains unaffected.

\subsection{Global \HI profiles}

\begin{figure}
\centering
\includegraphics[width=12cm, angle=0]{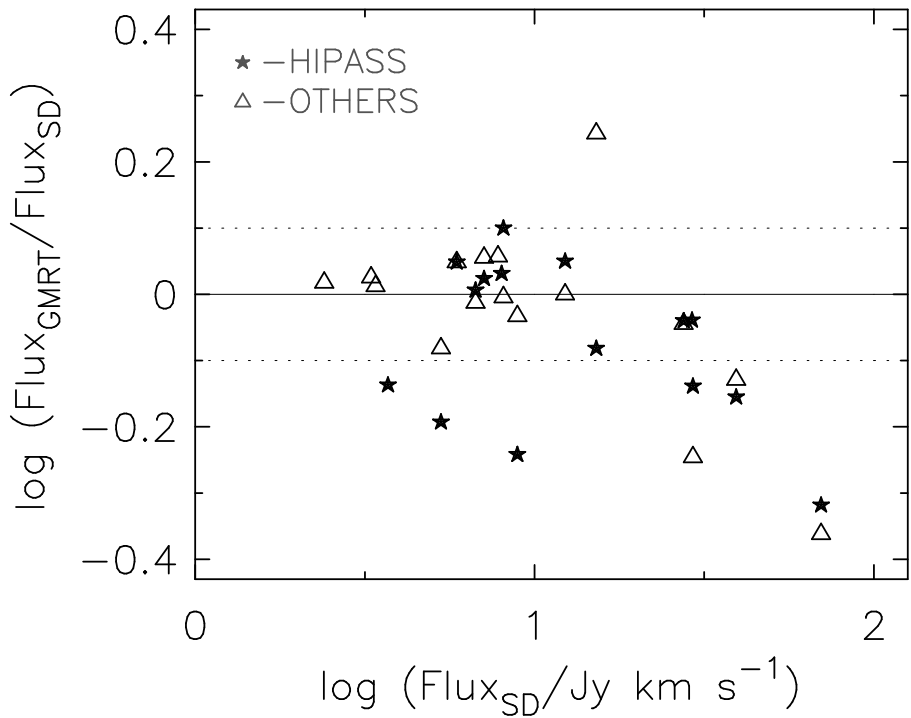}

\caption{A comparison of the integrated \HI flux densities of Eridanus galaxies from the
GMRT with those from the single dish data published elsewhere and from the HIPASS. Most
of the ratios are within $\pm$25\% of unity (indicated by the dotted lines).}

\label{fig:HIfluxComp}
\end{figure}

The integrated \HI flux density as a function of velocity is the global \HI profile as
would have been obtained from a single dish observation. The low resolution ($50''$)
cubes were used to obtain the global profiles as these cubes are most sensitive to the
diffuse emission.  The lower resolution zeroth order moment map was used to mark the
region over which the flux density was estimated in the channel images. The \HI mass is
obtained by using the following relation:

\begin{equation}
M_{\HI} (M_{\odot}) = 2.36\times10^{5} D^{2} ~{\delta}v \sum_{j=1}^{N_{chan}} S_{j}  
\end{equation}

\noindent where D is the distance in Mpc, $S_{j}$ is the integrated flux in Jy in the
spectral channel $j$ of velocity width ${\delta}v$ in \km. The distance is taken as
23~Mpc. In Fig.~\ref{fig:HIfluxComp}, the integrated \HI flux (${\delta}v
\sum_{j=1}^{N_{chan}} S_{j}$) from the GMRT is compared  with that from the HIPASS data
and from other single dish data. Some of the galaxies with higher values of the
integrated \HI flux in the single dish data show significantly less flux in the GMRT. The
\HI disk sizes of these galaxies are among the largest ($>6'$) in our sample. We believe
that the loss of flux for the large galaxies in the GMRT images is due to inadequate sampling
of shorter $(u,v)$ spacings in the GMRT.

\subsection{\HI line-width}

The \HI global profiles often peak at the two extreme ends of the rotation velocities of 
galaxies. The detailed shape of an \HI profile depends on the rotation curve, the
inclination and the \HI distribution in the galaxy. The \HI line-widths are broadened
from their true values due to random motions in the \HI gas and due to the finite spectral
resolution. The \HI line-width is a crucial parameter for studying the Tully-Fisher (TF)
relation. Most of the TF studies still use the \HI line-widths obtained from single dish
observations because of the simplicity of such observations.  The synthesis data of the
Eridanus group of galaxies provide an opportunity to compare the corrected \HI
line-widths obtained from the single dish \HI profiles with those obtained from the \HI
rotation curves. 

For those cases in which double-peaked \HI profiles are seen, \HI line-widths were estimated
at 20\% (W$_{20}$) and at 50\% (W$_{50}$) levels of the peak intensities at the two ends of
the \HI profiles. The locations of the two peak intensities were estimated separately
using Gaussian fits to the profiles. Bottinelli et al. (1990) derived an empirical
relation to correct for the instrumental broadening. They convolved the \HI profile
progressively with coarser velocity resolutions for a model galaxy, and determined the
broadening. A linear relationship between the channel resolution and the instrumental
broadening was suggested.  The broadening correction is estimated as $\delta W = 0.55
\times {\delta}V_{i}$ for W$_{20}$ and $\delta W = 0.13 \times {\delta}V_{i} $ for
W$_{50}$ for an instrumental resolution of ${\delta}V_{i}$. For the current observations,
${\delta}V_{i} = 13$~\km, implying that the corrections are $\sim7$~\km for  W$_{20}$ and
$\sim2$~\km for  W$_{50}$. 

A linear summation of the rotation velocity and the random velocity is appropriate to
estimate the observed width for the cases where the intrinsic width is almost boxy (i.e.,
in fast rotating galaxies). However, a summation in quadrature will be required for the slow
rotating (e.g., dwarf) galaxies where the solid body rotation together with the radial
distribution of the \HI gas will lead to an almost Gaussian profile. A composite relation
for all galaxies was given by Tully \& Fouque (1985). According to their relation, the
width due to the rotation motion $W_{R}$, the width due to random and turbulent motions 
$W_{t}$, and the observed width $W_{l}$ are related by:

\begin{equation}
W_{l}^{2} = W_{R,l}^{2}  - W_{t,l}^{2} \left( 1 - 2 e^{-(W_{l}/W_{c,l})^{2}}\right) + 2 W_{l} W_{t,l}
\left(1 - e^{-(W_{l}/W_{c,l})^{2}}\right)
\label{equ:width}
\end{equation}

\noindent where the subscript $l$ refers to the level (20\%, or 50\%) at which the widths
are estimated. The $W_{t,l}$ is estimated as $2 k_{l} \sigma$ from the velocity
dispersion of the \HI gas ($\sigma$) due to random and turbulent motions. The constant
factor $k_{l}$ is 1.80 at 20\% level and 1.18 for the 50\% level for a Gaussian profile.
The value of $\sigma$ is taken as 6~\km. $W_{c,l}$ is a parameter which defines the
transition region from linear to quadratic sum. The eqn.~\ref{equ:width} does a linear
subtraction if $W_{l} > W_{c,l}$ and a quadratic subtraction if $W_{l} < W_{c,l}$. The
values of $W_{c,l}$ are determined empirically as 120~\km for the 20\% level and 100~\km
for the 50\% level by Tully \& Fouque (1985).

\begin{figure}
\centering
\includegraphics[width=12cm, angle=0]{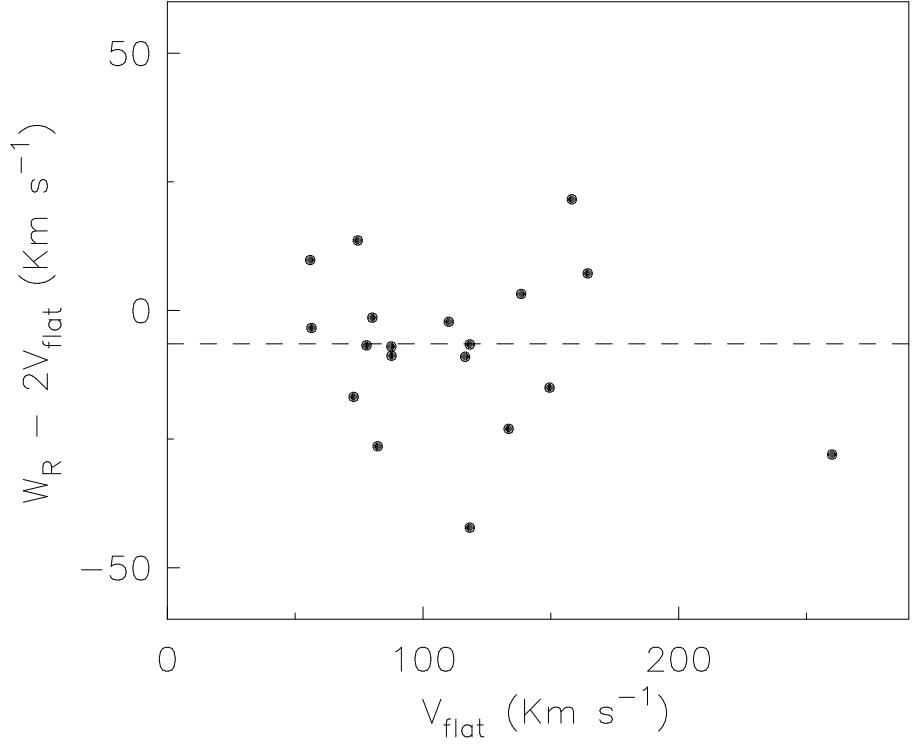}
\caption{Comparison of the corrected \HI widths ($W_{R;50}$) with the flat rotation velocities of the
Eridanus galaxies. The dashed line is the mean value at $6.5$~\km.}
\label{fig:WVrot}
\end{figure}

The corrected \HI widths are compared with the flat rotation velocities of the galaxies
in Fig.~\ref{fig:WVrot}. The mean value of the difference ($W_{R,50}-2V_{flat}$) is
$\sim6.5$~\km. 

\subsection{Rotation curves}

The rotation curves were derived using the tilted ring model (cf. Begeman 1989).  The
{\sf GIPSY} task {\sf ROTCUR} was used. The basic methodology of this model is the
following. The model assumes the gas to be in circular orbits. The position angle and the
inclination of the \HI disk are allowed to vary with radius. The fitting procedure
generally involves estimation of 5 unknowns, $viz.$, the dynamical centre (X,Y), the
systemic velocity (V$_{sys}$), the position angle (PA) of the major axis, the inclination
angle (Incl), the circular rotation velocity (V$_{rot}$), and optionally the expansion
velocity $V_{exp}$. The observed radial velocity V(x,y) at a rectangular sky coordinate
(x,y) and at a radius $r$ is given by - 

\begin{equation}
V(x,y) \&= V_{sys} + V_{rot}~cos(\theta) sin(INCL) + V_{exp}~sin(\theta) sin(INCL)
\end{equation} 
where\\
\begin{equation}
cos(\theta) \&= \frac{-(x-X)sin(PA) + (y-Y)cos(PA)}{r}
\end{equation}
\begin{equation}
sin(\theta) \&= \frac{-(x-X)cos(PA) + (y-Y)sin(PA)}{r~cos(INCL)}
\end{equation}

The $V_{exp}$ term was not fitted in the present analysis, and was kept fixed at zero.
$V_{exp}$ can be used to estimate the non-axisymmetry in the velocity field. The velocity
fields at  $20''$ resolution obtained via Gaussian fits were used in this analysis. The
iterative procedure described by Begeman (1989) is used to estimate the disk's
kinematical parameters and the rotation curve. The velocities are averaged in elliptical
annuli of width $10''$ and extracted as a function of the azimuth at different
radii in increments of $10''$.  The velocity fields in the receding side and in the
approaching side were fitted separately to obtain two rotation curves. The iterative
scheme to estimate the different parameters is described below.

\subsubsection{Dynamical centre (X,Y) and systemic velocity ($V_{sys}$)}

The velocity determined from the \HI width or the optical velocity of the
galaxy was used as the initial guess for the systemic velocity. The guesses for the
centre, the position angle and the inclination angle were their respective estimates
obtained from the ellipse fits to the optical isophotes. The first iteration is started
by fixing the inclination and the position angle, and fitting the centre and the systemic
velocity.  If the velocity field is symmetric, the fitted values of the dynamical centre
and the systemic velocity should be similar for all the rings. However, often galaxies do
not have symmetric velocity fields due to kinematical lopsidedness. Therefore, slightly
different estimates of centre and systemic velocity may result from each ring.  An
overlay of the velocity field contours over the optical image helps in deciding the
quality of the fit.   For a galaxy with no warp, the velocity field lines should run
straight along the minor axis. The line joining the cusps of the iso-velocity contours of
identical rotation velocity at the two halves of the galaxy should trace the direction of
the major axis. The intercept of the major axis with the minor axis is expected to be the
dynamical centre.  

If a galaxy has a warp, the velocity field lines will show a characteristic integral sign
shaped structure. Often by a visual inspection of the velocity field, it was possible to
decided whether reasonably good fits were obtained or not. It was found in some cases
that the dynamical centre was not coincident with the optical centre (e.g., in IC~1953).
The centre and the systemic velocity were computed as un-weighted mean of their values in
all the rings for which satisfactory solutions could be obtained. Once the centre and the
systemic velocity are determined, these values are held fixed for all radii during
successive iterations.

\subsubsection{Position angle}

In this step, the position angle is determined as a function of radius. The inclination
angle for each radius is kept fixed and Vrot is allowed to vary. The variations in
position angle with radii can be inferred from a visual inspection of the velocity field.
If a galaxy is warped, the position angle varies gradually starting at  certain radii and
often becomes constant at large radii. A warp often results in variations in both the
inclination and the position angle simultaneously.  For  galaxies, where no significant
change in the position angle or in the inclination or in both were seen, an average value
of the position angle was taken. 

\subsubsection{Inclination}

In this iteration, the variations in the inclination angle are modeled as a function of
radius. The previously determined parameters (centre, systemic velocity, and PA) are kept
fixed. Reliable estimates for the inclination can not be made for galaxies having low
inclination (usually below $45\deg$),  using the tilted ring model (Begeman 1989). In
such cases, optical inclination angles are more reliable.

\subsubsection{Rotation velocities}

In the final iteration, all the disk parameters obtained from the previous iterations are
kept fixed, and a fit is carried out to obtain the circular rotation velocity as a
function of radius. The receding and the approaching sides of the galaxies were fitted
separately.  In this step, a cosine weighting scheme was adopted such that points along
the major axis have maximum weight and points along the  minor axis have zero weight. In
addition, points within $\pm$20 degree from the minor axis were excluded. This is due to
the fact that the minor axis does not have any information on the rotation velocity, and
the points along the major axis give a direct estimate of the projected rotation
velocity. 

The rotation velocities obtained from the tilted ring fit should be corrected for the
effects of beam smearing. The effects of  beam smearing are maximum at the locations of
steep velocity gradients. The flat portion of the rotation curve remains nearly
unaffected due to beam smearing. The rotation curves presented here are not corrected for
the effects of beam smearing. Therefore, for those analyses which are critically dependent on the
quality of the rotation curves (e.g., mass modeling), these rotation curves should not
be used without making an appropriate correction. Such analyses will be presented
elsewhere.

\subsection{Radial \HI density profiles}

The total \HI maps were used to estimate the mean radial \HI surface density profiles by
azimuthally averaging the \HI column densities in concentric elliptical annuli. The axial
parameters of the ellipses were obtained from the tilted ring models. The elliptical
annuli were sampled at intervals of $10''$ to get  two points in one synthesised beam of
$20''$.  The widths of the annuli were kept fixed at $10''$. The average radial profile
in each annulus is scaled by the ratio of the total \HI mass to the summed column density
in all annuli to obtain the \HI mass surface density. The profiles are corrected for the
projection effects to obtain the face-on mass surface density in units of
M$_\odot$~pc$^{-2}$. The profiles can  be quite uncertain in high inclination galaxies
where some flux density from the lower radii along the minor axis will be included at
larger radii due to finite spatial resolution. Also, the profiles will artificially
extend to larger radii near the outer edge  of the \HI disk due to the finite angular
resolution.

\subsection{Dynamical mass} 

The dynamical mass can be estimated from the rotation curves. The total mass within a
radius $R$ (kpc)  can be derived using the relation $M_{tot} (M_{\odot}) = 2.3 \times
10^{5} V_{rot}^{2} R/G$, where   $V_{rot}$ is the rotation velocity in \km. There can be
several choices for the radius $R$, e.g, \HI disk radius, optical disk radius, disk scale
length etc. The dynamical masses were estimated within the optical radius (i.e.
$D_{25}/2$) in the present analysis. The estimates were made only for those galaxies in
which flat rotation curves were detected.

\section{\HI properties of galaxies}

In this section, some of the \HI properties of galaxies derived from the GMRT
observations are presented and are compared with those for nearby galaxies in field and
in other loose groups.

\subsection {\HI and total dynamical mass}

\begin{figure}
\centering
\includegraphics[width=8cm, angle=0]{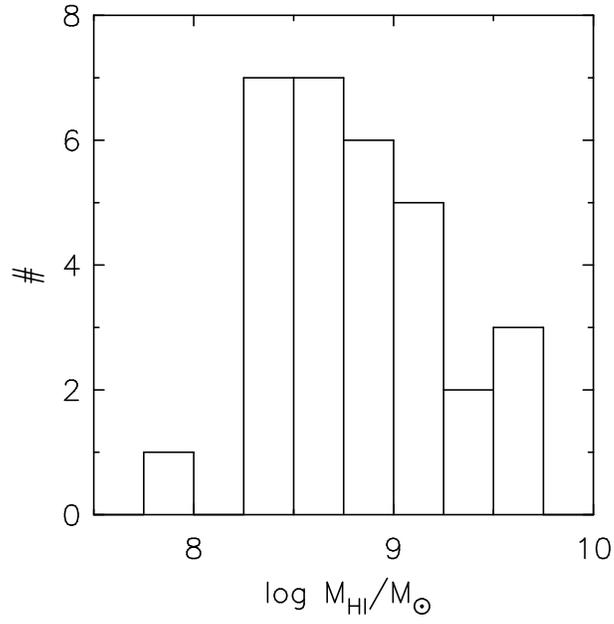}
\caption{The histogram of \HI masses of the Eridanus galaxies observed with the GMRT.}
\label{fig:hi-hist}
\end{figure}

The histogram of \HI masses of galaxies detected by GMRT is shown in
Fig.~\ref{fig:hi-hist}. The lowest \HI mass ($\sim8\times10^{8}$~M$_{\odot}$) detected
is that of an S0 galaxy NGC~1481. Some galaxies toward the high mass end in the histogram
will have their \HI masses underestimated by GMRT. Due to the limited sample size, this
plot is of limited statistical significance to estimate the \HI mass function.

\begin{figure}
\centering
\includegraphics[width=8cm, angle=0]{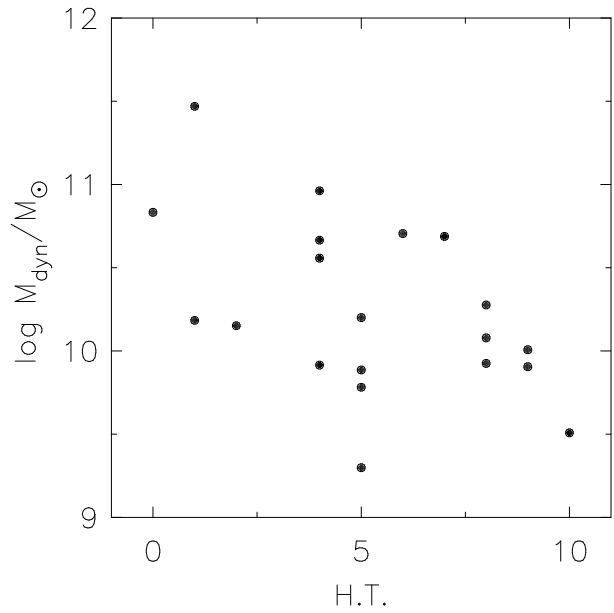}
\caption{Dynamical mass plotted against the Hubble type. The masses are estimated within the optical
diameter (D$_{25}$).}
\label{fig:dyna}
\end{figure}

The dynamical masses are plotted in Fig.~\ref{fig:dyna} as a function of Hubble type. Only
those galaxies are plotted whose flat part of the rotation curve could be measured. There seems
to be a systematic trend in the sense that early type galaxies have on average higher dynamical
masses than the late type galaxies. This result is consistent with other studies (e.g., Broeils
\& Rhee 1997, Verheijen \& Sancisi 2001). 

The statistical significance of the correlation was estimated using the Spearman Rank-Order Correlation
Coefficient method. The significance of the non-zero correlation coefficient in Fig.~\ref{fig:dyna}
is $\sim97$\%.

\subsection{\HI mass to luminosity ratio}

\begin{figure}
\centering
\includegraphics[width=12cm, angle=0]{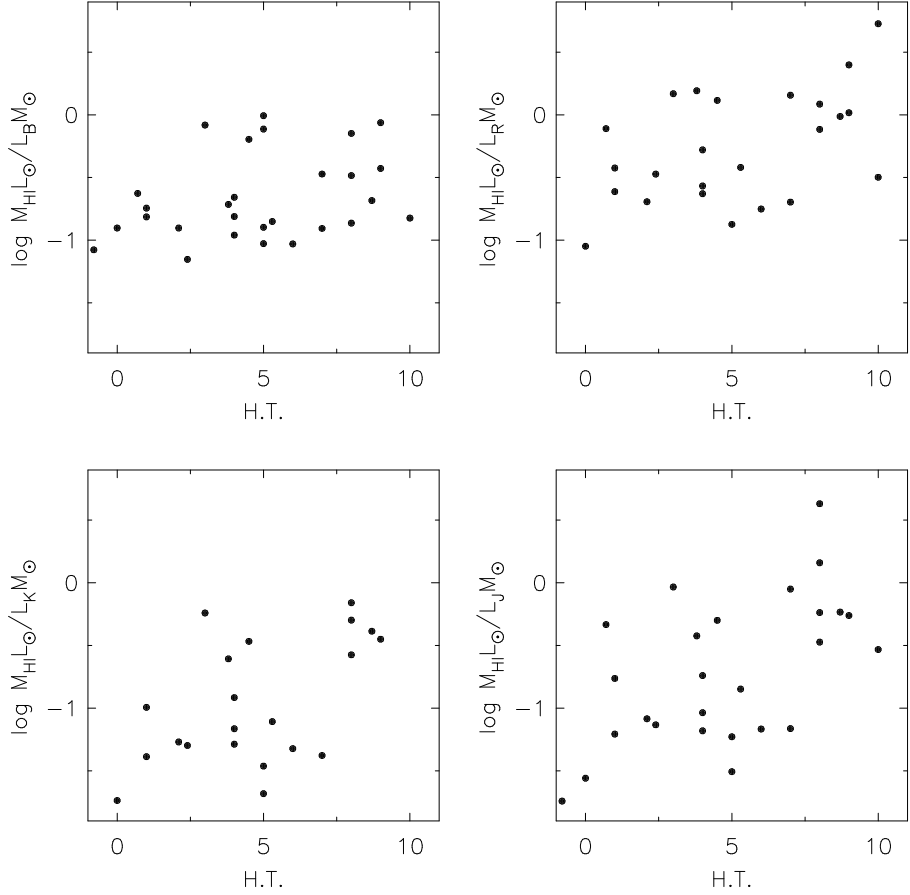}

\caption{The \HI mass to luminosity ratio in the optical and the near-IR bands. There
appears to be a trend in the R, and in the near-IR bands in the sense that late
type galaxies have on an average higher value of $M_{\HI}/L$.}

\label{fig:MtoL}
\end{figure}

The ratio of the \HI mass to the luminosities in the B, R, J, and K bands are plotted in
the different panels of Fig~\ref{fig:MtoL} respectively against the Hubble types.  
The B-band luminosities are estimated using
the face-on magnitudes from the RC3 catalog, and assuming the Sun's absolute magnitude in
the B-band to be 5.48. The values of $M_{\HI}/L_{B}$ are in general consistent with that
of Broeils \& Rhee (1997). No significant trend in the $M_{\HI}/L$ ratio with respect to
the type is seen in the B-band. The significance of any correlation is less than
$\sim70$\%.

However, significantly higher probabilities ($>97$\%) were found for a correlation between 
$M_{\HI}/L$ and H.T. in the other bands. The correlations in the R and the near-IR (J \& K) bands
are in the sense that late type galaxies have on an average higher $M_{\HI}/L$ value. This trend
in $M_{\HI}/L$ in the J and K bands for which the extinction corrections are relatively less
significant are  consistent with the results for the Ursa-Major galaxies (Verheijen \& Sancisi
2001).

\vspace{0.1in}

\subsection{Ratio of  \HI diameter to  optical diameter}
 
\begin{figure}
\centering
\includegraphics[width=12cm, angle=0]{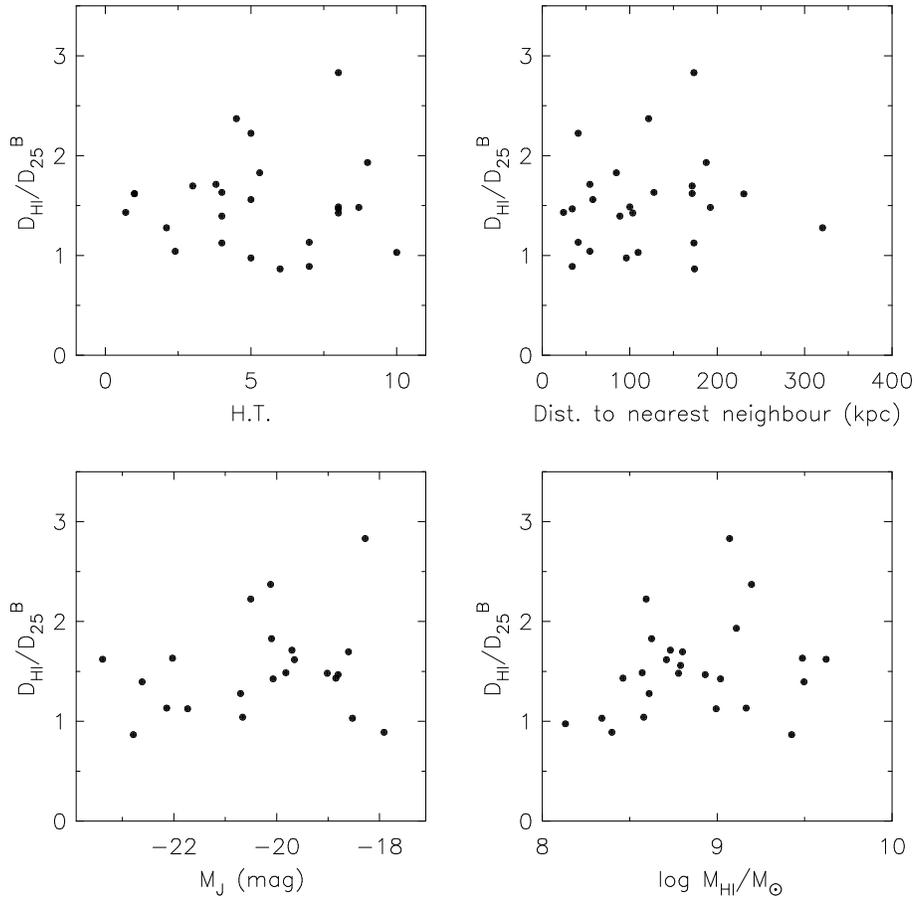}

\caption{The ratio of the \HI diameter to the optical diameter is plotted with
different properties of galaxies.  The mean ratio of the sample  is
$1.7\pm0.8$. This ratio is independent of the Hubble type, the \HI mass, the
distance to the nearest neighbor, and the J-band luminosity. }

\label{fig:DHItoD25}
\end{figure}

The ratio of the \HI diameter to the optical diameter is plotted in Fig.~\ref{fig:DHItoD25}
against various properties of the galaxies.   No significant correlation is seen with the
Hubble type, the \HI mass, the projected distance to the nearest neighbor, and the J-band
magnitude. On contrary to this, Verheijen \& Sancisi (2001) found a significant correlation
between $D_{\HI}/D_{25}$ and J-band magnitude as well as Hubble type for the Ursa-major group of
galaxies. The average value of $D_{\HI}/D_{25}$ is $1.7\pm0.8$, which is consistent with the
value of $1.7\pm0.5$ obtained by Broeils \& Rhee (1997) for a sample of nearby luminous
galaxies .

\begin{figure}
\centering
\includegraphics[width=10cm, angle=0]{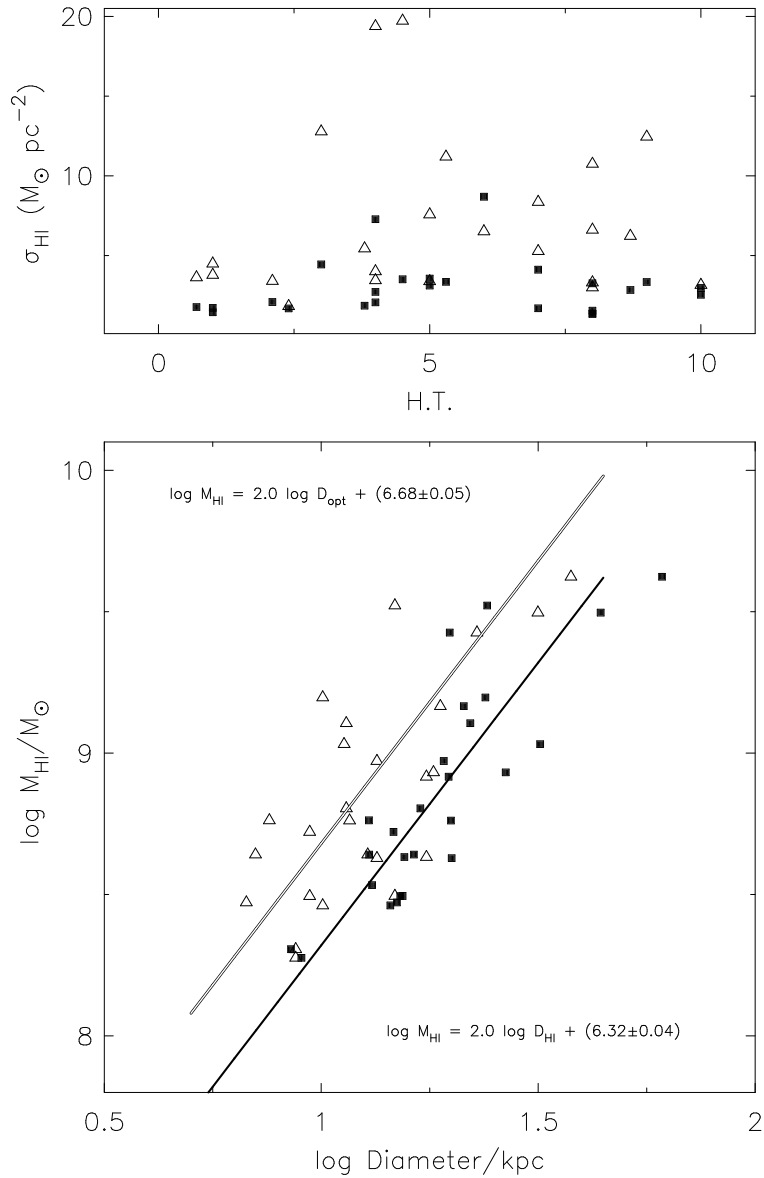}

\caption{(Lower panel) The \HI mass is plotted  against  \HI diameter (filled
squares) and optical diameter (open triangles).  The two lines are the best fits to
the two sets of the data respectively.  The slopes were kept fixed at 2.
(Upper panel) The \HI mass surface density estimated over the \HI disk (filled
squares) and over the optical disk (open triangles) plotted against the Hubble type.}

\label{fig:HIsurf}
\end{figure}

\vspace{0.1in}

\subsection{\HI surface density} 

The \HI masses of galaxies are observed to correlate with their \HI diameters and
optical diameters (Haynes \& Giovanelli 1984, Broeils \& Rhee 1997, Verheijen \& Sancisi
2001). The latter authors showed that the correlation is tighter with the \HI diameter
than with the optical diameter with a slope of $\sim2$. The lower panel of
Fig.~\ref{fig:HIsurf} shows  log M$_{\HI}$ for the Eridanus galaxies plotted against log
$D$ for both optical and \HI diameters. A straight line is fitted with a fixed slope of
2.  It can be seen that the \HI diameters of the galaxies in the Eridanus group are more
tightly correlated than the optical diameters with their \HI masses. The upper panel of
Fig.~\ref{fig:HIsurf} shows the distribution of the \HI mass surface densities
($\sigma_{\HI} = 4 M_{\HI}/\pi D^{2}$) of Eridanus galaxies  against their Hubble types.
It can be seen that the  mass surface densities over the \HI disks are less scattered
compared to the mass surface density over the optical disks. 

A summary of the
various estimated parameters of the Eridanus galaxies can be found in Table 5.

\section{Summary}

The Eridanus group of galaxies were observed in the \HI 21cm-line using the GMRT for
$\sim200$ hour. There is considerable scope to use this data for several studies which
will be presented in subsequent papers. The following conclusions can be drawn from the
\HI properties of the Eridanus galaxies:

\begin{itemize}

\item The early type disk galaxies seem to have higher dynamical masses
compared to the late type disk galaxies. 

\item The $M_{\HI}/L$ ratio shows a trend with respect to the type in the R-band
and in the near-IR (J \& K) bands in the sense that late type galaxies have on
an average higher $M_{\HI}/L$.

\item The average value of $D_{\HI}/D_{opt}$ is $1.7\pm0.8$ for the Eridanus galaxies is
consistent with that for galaxies in other groups and fields.

\item The ratio of $M_{\HI}/D_{\HI}$ has less scatter compared to the ratio of
$M_{\HI}/D_{opt}$ for different galaxy types.

\end{itemize}

\section*{Acknowledgments} 

AO would like to acknowledge the encouragement 
and the support of late K. R. Anantharamaiah in beginning this 
observational program with the GMRT.  
We thank Marc Verheijen, Jayaram Chengalur, and the referee for useful comments
on various issues discussed here.  We thank the staff of the GMRT who made these
observations possible. The GMRT is operated  by the National Centre for Radio
Astrophysics of the Tata Institute of Fundamental Research. This research has
been benefited by the NASA's Astrophysics Data System (ADS) and Extra-galactic
Database (NED) services. The open softwares used in this research are greatly
acknowledged.

\clearpage

\appendix

\section{Tables}

\vspace{0.1in}

The description of the columns in Tab.~\ref{tab:sample} are as follows: 

\vspace{0.1in} 
\noindent$Column~1:$ Serial number. \\
$Column~2:$ Name of the galaxy. \\ 
$Columns~3\&4:$ Equatorial coordinates in the J2000 epoch.\\ 
$Column~5\&6:$ Super Galactic coordinates.\\ 
$Column~7:$ Helio-centric velocity (optical definition). 

\vspace{0.1in}

The description of the columns in Tab.~\ref{tab:prop} are as follows:

\vspace{0.1in}

\noindent$Column~1:$ Serial number. \\
$Column~2:$ Name of the galaxy. \\
$Column~3:$ Morphological type. \\
$Column~4:$ Hubble type. \\
$Column~5:$ Apparent total photographic B-band magnitude corrected for inclination. \\
$Column~6:$ Apparent total K-band magnitude from 2MASS. \\
$Column~7:$ B--K color. \\
$Columns~8\&9:$ Single dish \HI flux integral (Jy~\km) and velocity width (\km) at 20\% of the peak flux 
density in the global \HI profile.

\vspace{0.1in}

The description of the columns in Tab.~\ref{tab:Radobs} are as follows:

\vspace{0.1in}

\noindent$Column~1:$ Serial number.\\
$Column~2:$ Date of observation. \\
$Column~3:$ Coordinates of the pointing centre. \\
$Column~4:$ Galaxies in the field of view. \\
$Column~5:$ Separation (arc min) of galaxy(ies) in the field of view from the pointing centre. \\
$Column~6:$ RMS noise (mJy beam$^{-1}$) in the channel images. \\

\vspace{0.1in}

The results are presented in Tab.~\ref{tab:results}. The description of the entries in the table is
the following: 

\vspace{0.1in}

\noindent{\it Column~1} -- {\sf Galaxy name, Opt. Centre, Radio Centre:} The abbreviated names of
galaxies. The optical centre is from the RC3 catalog and the radio centre is estimated  from the
tilted ring fit. \\
{\it Column~2} -- {\sf Type, INCL, PA:} The morphological type, the inclination, and the position
angle.\\
{\it Column~3} -- {\sf Velocities:} The systemic velocities (Helio-centric) obtained from the tilted
ring fit, from the global profile fit, and from RC3. \\
{\it Column~4} -- {\sf  Diameters:} The \HI diameter, the B-band optical diameter at B 25
mag~arc sec$^{-2}$ from RC3, and the R-band optical diameter at R 25 mag~arc sec$^{-2}$ from the 
current observations.\\
{\it Column~5} -- {\sf Rotation velocities:} The entries in this column are maximum rot. velocity
from the rotation curve, velocity at the flat part of the rotation curve, and maximum rot. velocity
estimated from the \HI global profile corrected for the broadening due to channel width and random
and turbulent motions in galaxies.\\
{\it Column~6} -- {\sf Masses:} The \HI mass, the stellar mass from the K-band light assuming
$M^{*}/L$ ratio of 0.6, and the dynamical mass within B25.\\
{\it Column~7} -- {\sf Mass to light ratios:} The ratios of the \HI mass to the B-band, the R-band, and
the K-band luminosities.\\
{\it Column~8} -- {\sf Mean \HI surface densities and the ratio of the \HI diameter to the optical
diameter:} The mean \HI surface density (M$_{\odot}$/pc$^{2}$) over the \HI diameter and the optical
diameter at B25 mag~arc sec$^{-2}$. The last entry in this column is the ratio of the \HI diameter
to the optical diameter at 25 mag~arc sec$^{-2}$.

\section{\HI atlas}

\begin{figure} 
\centering 
\includegraphics[width=11cm, angle=0]{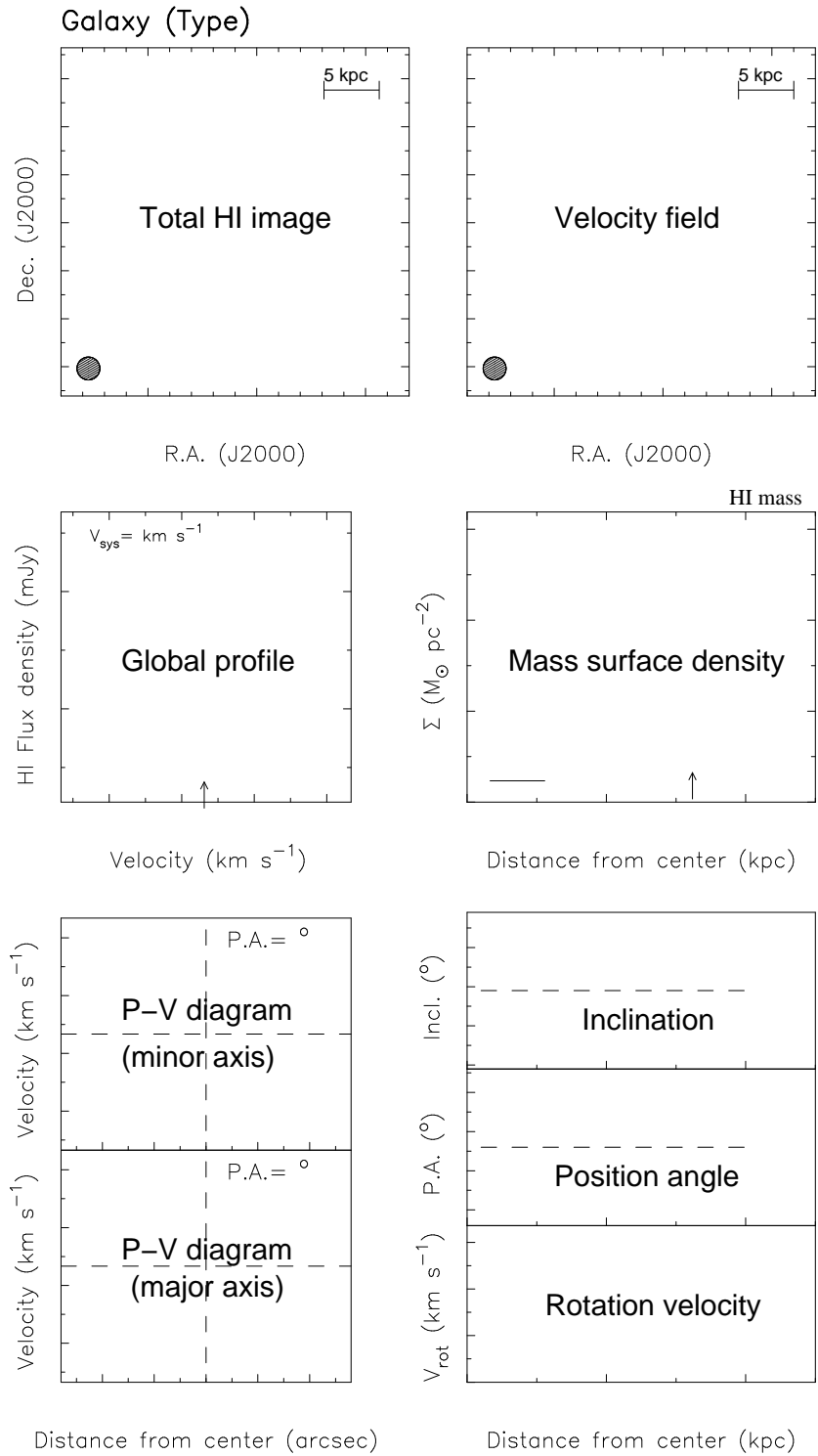}
\caption{The layout of the atlas}
\label{fig:layout}
\end{figure}

The results are also presented in the form of an atlas. The layout of the atlas is shown
in Fig.~\ref{fig:layout}.  The high resolution images ($20''-30''$) are presented in this
atlas. The contents of the panels in the atlas are described below: 

{\sf Total \HI image}\\ Contours of  total \HI  image are overlaid upon a gray-scale
image of the galaxy from {\sf DSS} (Digitized Sky Survey). The contours start at $N_{\HI}
= 10^{20}$ cm$^{-2}$ with increments of $2\times10^{20}$ cm$^{-2}$. The first contour is
white in color and rest of them are black. The contour lines become thicker (appearing
darker) in two steps for higher \HI column densities. One step is at $N_{\HI} =
1.2\times10^{21}$ cm$^{-2}$ till $N_{\HI} = 2\times10^{21}$ cm$^{-2}$ and the second step
is for higher column densities.  The synthesised beam is shown at the bottom left hand
corner. A horizontal bar of length 5 kpc is drawn near the upper right hand corner. 

{\sf Velocity field}\\ The velocity field of the galaxy is shown as iso-velocity contours
overlaid upon the gray-scale optical DSS image. The white contours indicate velocities in
the approaching half and dark contours indicate velocities in the receding half of the
galaxy. The contours are drawn in increments of 10 \km. The thick dark contour near the
centre of the galaxy represents the systemic velocity of the galaxy estimated from the
tilted ring model. A horizontal bar of length 5 kpc is drawn near the upper right hand corner.

{\sf \HI Global profile} \\  The \HI global profile is shown as a function of  the
Heliocentric velocity (optical definition). The systemic velocity of the galaxy estimated
from the tilted ring model is written near the top left hand corner of the panel, and
also is marked by an arrow along the velocity axis.

{\sf \HI mass surface density} \\ The face-on \HI mass surface density is shown in this
panel. The values in the approaching and the receding halves are indicated by open and
filled circles respectively. The average profile is shown by a thick line.  An arrow is
marked along the abscissa at the position corresponding to the optical radius estimated
at the 25 mag~arc sec$^{-2}$ B-band isophotal level from the RC3 catalog. A horizontal bar
at the left bottom indicates the spatial resolution of the total \HI image.

{\sf Position-Velocity diagrams} \\ The Position-Velocity (PV) diagrams are shown along
the major and the minor axes of the galaxy in the bottom and the top panels respectively.
The contours are from 3~mJy~beam$^{-1}$ to 8 mJy~beam$^{-1}$ in increments of
1~mJy~beam$^{-1}$, and thereafter, in increments of 4~mJy~beam$^{-1}$. A  vertical
dotted line marks the dynamical centre and a horizontal dotted line marks the systemic
velocity, both determined from  the tilted ring model. The velocity resolution and the
spatial resolution are indicated by the vertical and the horizontal scale-bar
respectively at the bottom left hand corners of the two diagrams. 

{\sf Axial parameters and rotation curve} \\ The axial parameters (PA and Incl)
determined from the tilted ring models are plotted in the upper two panels of this
figure. The thin lines in these two panels show the final adopted trend of the variations
in PA and Incl respectively. The values of the position angle and the inclination
determined from the optical images are also marked in the respective panels by an
asterisk sign. The rotation curves are shown in the bottom most panel. The approaching
and the receding halves are plotted separately using open and filled symbols
respectively. The average rotation curve is drawn by a thin line. These rotation curves 
are not corrected for the effects of beam smearing.

\begin{table}
\begin{center}
\caption{Sample of galaxies}
\label{tab:sample}
\begin{tabular}{rlccccc}
\hline
\hline
\bf{\#} & \bf{Name} 	& \bf{$\alpha$}\small(J2000) & \bf{$\delta$}\small(J2000) &
\bf{$l_{SG}$} & \bf{$b_{SG}$} & \bf{$cz$}\\
	 & 	     	& ~h~~m~~s~ & ~~\hbox{$^\circ$}~~$'$~~$''$~ & \hbox{$^\circ$} &
	 \hbox{$^\circ$} & (km/s) \\
\hline \\

 1 & NGC~1297 		& 03~19~14.2 & --19~06~00 & 284.30 & --38.61 & 1579 \\
 2 & NGC~1309		& 03~22~06.5 & --15~24~00 & 289.12 & --38.86 & 2135 \\
 3 & NGC~1315 		& 03~23~06.6 & --21~22~31 & 281.46 & --39.68 & 1673 \\
 4 & SGC 0321.2-1929 	& 03~23~25.1 & --19~17~00 & 284.18 & --39.61 & 1545 \\
 5 & UGCA~068	 	& 03~23~47.2 & --19~45~15 & 283.58 & --39.73 & 1838 \\
 6 & NGC~1325 		& 03~24~25.4 & --21~32~36 & 281.26 & --39.99 & 1589 \\
 7 & ESO~548-~G~016	& 03~26~02.4 & --21~20~26 & 281.55 & --40.36 & 2119 \\
 8 & NGC~1332 		& 03~26~17.3 & --21~20~07 & 281.56 & --40.41 & 1524 \\
 9 & NGC~1331		& 03~26~28.3 & --21~21~20 & 281.54 & --40.46 & 1210 \\ 
10 & APMBGC~548+070+070 & 03~26~31.3 & --21~13~01 & 281.72 & --40.46 & 1548 \\
11 & ESO~548-~G~021     & 03~27~35.3 & --21~13~42 & 281.72 & --40.71 & 1668 \\
12 & ESO~548-~G~025	& 03~29~00.7 & --22~08~45 & 280.53 & --41.08 & 1680 \\
13 & NGC~1345		& 03~29~31.7 & --17~46~40 & 286.32 & --40.91 & 1529 \\
14 & NGC~1347		& 03~29~41.8 & --22~16~45 & 280.35 & --41.24 & 1759 \\
15 & NGC~1353		& 03~32~03.0 & --20~49~09 & 282.34 & --41.73 & 1525 \\
16 & UGCA 077		& 03~32~19.2 & --17~43~05 & 286.48 & --41.57 & 1961 \\
17 & ESO~482-~G~005	& 03~33~02.2 & --24~07~58 & 277.89 & --42.03 & 1915 \\
18 & IC~1952		& 03~33~26.7 & --23~42~46 & 278.46 & --42.13 & 1812 \\
19 & ESO~548-~G~036	& 03~33~27.6 & --21~33~53 & 281.35 & --42.09 & 1480 \\
20 & IC~1953		& 03~33~41.9 & --21~28~43 & 281.47 & --42.14 & 1867 \\
21 & NGC~1359		& 03~33~47.7 & --19~29~31 & 284.15 & --42.06 & 1966 \\
22 & ESO~548-~G~043	& 03~34~10.5 & --19~33~30 & 284.07 & --42.16 & 1931 \\
23 & ESO~548-~G~044	& 03~34~19.2 & --19~25~28 & 284.25 & --42.18 & 1696 \\
24 & NGC~1371		& 03~35~02.0 & --24~55~59 & 276.80 & --42.48 & 1471 \\
25 & NGC~1370		& 03~35~14.6 & --20~22~25 & 282.99 & --42.46 & 1063 \\
26 & ESO~548-~G~049	& 03~35~28.1 & --21~13~01 & 281.85 & --42.55 & 1510 \\
27 & IC~1962		& 03~35~37.5 & --21~17~39 & 281.74 & --42.58 & 1806 \\
28 & NGC~1377		& 03~36~39.1 & --20~54~08 & 282.29 & --42.81 & 1792 \\
29 & ESO~482-~G~013	& 03~36~53.9 & --24~54~46 & 276.82 & --42.90 & 1835 \\
30 & NGC~1385		& 03~37~28.3 & --24~30~05 & 277.38 & --43.04 & 1493 \\
31 & NGC~1383		& 03~37~39.2 & --18~20~22 & 285.80 & --42.89 & 1948 \\
32 & NGC~1390		& 03~37~52.2 & --19~00~30 & 284.89 & --42.99 & 1207 \\
33 & NGC~1393		& 03~38~38.6 & --18~25~41 & 285.70 & --43.13 & 2185 \\
34 & NGC~1401		& 03~39~21.8 & --22~43~29 & 279.81 & --43.48 & 1495 \\
35 & ESO~548-~G~064	& 03~40~00.0 & --19~25~35 & 284.36 & --43.52 & 1694 \\
36 & ESO~548-~G~065	& 03~40~02.7 & --19~22~00 & 284.45 & --43.53 & 1221 \\
37 & IC~0343		& 03~40~07.1 & --18~26~36 & 285.72 & --43.48 & 1841 \\
38 & NGC~1407		& 03~40~11.9 & --18~34~49 & 285.53 & --43.51 & 1779 \\
39 & ESO~482-~G~031	& 03~40~41.5 & --22~39~04 & 279.92 & --43.79 & 1621 \\
40 &APMBGC~548-110-078  & 03~40~52.5 & --18~28~29 & 285.69 & --43.66 & 1595 \\
41 & NGC~1415		& 03~40~56.7 & --22~33~47 & 280.04 & --43.85 & 1585 \\
42 & NGC~1414		& 03~40~57.0 & --21~42~47 & 281.22 & --43.84 & 1681 \\
43 & ESO~548-~G~072	& 03~41~00.3 & --19~27~19 & 284.34 & --43.76 & 2034 \\
44 & NGC~1416		& 03~41~02.9 & --22~43~09 & 279.82 & --43.87 & 2167 \\
45 & IC~0345		& 03~41~09.1 & --18~18~51 & 285.92 & --43.71 & 1335 \\
 
\end{tabular} 
\end{center} 
\end{table}

\begin{table}
\begin{center}
\begin{tabular}{rlccccc}
\hline
\hline
\bf{\#} & \bf{Name} 	& \bf{$\alpha$}\small(J2000) & \bf{$\delta$}\small(J2000) &
\bf{$l_{SG}$} & \bf{$b_{SG}$} & \bf{$cz$}\\
	 & 	     	& ~h~~m~~s~ & ~~\hbox{$^\circ$}~~$'$~~$''$~ & \hbox{$^\circ$} &
	 \hbox{$^\circ$} & (km/s) \\
\hline \\
46 & ESO~482-~G~035	& 03~41~15.0 & --23~50~10 & 278.27 & --43.91 & 1890 \\
47 & NGC~1422		& 03~41~31.1 & --21~40~54 & 281.27 & --43.97 & 1637 \\
48 & IC~0346		& 03~41~44.6 & --18~16~01 & 286.00 & --43.85 & 2013 \\
49 & ESO~549-~G~002	& 03~42~57.3 & --19~01~12 & 284.99 & --44.19 & 1111 \\
50 & MCG~-03-10-041	& 03~43~35.5 & --16~00~52 & 289.18 & --44.07 & 1215 \\
51 & NGC~1440		& 03~45~02.9 & --18~15~58 & 286.09 & --44.63 & 1534 \\
52 & NGC~1438		& 03~45~17.2 & --23~00~09 & 279.43 & --44.85 & 1555 \\
53 & NGC~1452		& 03~45~22.3 & --18~38~01 & 285.58 & --44.73 & 1737 \\
54 & ESO 549- G 018	& 03~48~14.1 & --21~28~28 & 281.60 & --45.52 & 1587 \\
55 & NGC~1481		& 03~54~28.9 & --20~25~38 & 283.15 & --46.96 & 1733 \\
56 & NGC~1482		& 03~54~39.3 & --20~30~09 & 283.04 & --47.00 & 1916 \\
57 & ESO~549-~G~035    	& 03~55~04.0 & --20~23~01 & 283.22 & --47.10 & 1778 \\
\hline \\
\hline \\
\multicolumn{7}{p{5.5in}}{\ch Note: The positions and velocities are from the {\it NASA Extra-galactic
Database (NED)}}
\vspace{5.8in}
\end{tabular} 
\end{center} 
\end{table}

\newpage

\begin{table}
\begin{center}
\caption{Optical \& \HI properties of galaxies}
\label{tab:prop}
\begin{tabular}{rllrrrrcr}
\hline
\hline
\bf{\ch \#} & \bf{\ch Name} & \bf{\ch Morph.} & \bf{\ch H.T.} & \bf{\ch B$_{0}^{T}$} & \bf{\ch K$^{T}$} & 
\bf{\ch B -- K} & \bf{\ch \HI Flux} & \bf{\ch W$_{20}$}\\ 
 & & & & (\ch mag) & (\ch mag) & (\ch mag) & (\ch Jy~km/s) & (\ch km/s) \\
\hline
%ESO~000-~G~000 & Sa & 000.0 & 00.0 & 000 & 00.0 & 00.0 & 00.0 \\
 1 & NGC~1297           &S0a  & --2.3 & 12.65 &  8.93 & 3.72 &      &\\
 2 & NGC~1309		&Sbc  &  4.0  & 11.83 &  9.10 & 2.73 & 18.7 & 171\\
 3 & NGC~1315           &S0   & --1.0 & 13.38 &  9.73 & 3.65 &      &\\
 4 & SGC 0321.2-1929	&Im   & 10.0  &       &       &      &  3.1 & 79\\
 5 & UGCA 068	        &Scdm &  8.7  & 13.56 &       &      &  5.9 & 131\\
 6 & NGC~1325           &Sbc  &  4.0  & 11.51 &  8.63 & 2.88 & 24.4 & 348\\
 7 & ESO~548-~G~016	&S?   &       & 14.61 &       &      &      &\\
 8 & NGC~1332           &S0   & --3.0 & 11.21 &  7.05 & 4.16 &      &\\
 9 & NGC~1331           &E/S0 & --5.0 & 14.22 & 10.76 & 3.46 &      &\\
10 & APMBGC~548+070+070 &S0   &       &       & 11.23 &      &      &\\
11 & ESO 548- G 021     &Sdm  &       &       &       &      &      &\\
12 & ESO 548- G 025	&Sa   &  0.7  & 14.47 &       &      &      &\\
13 & NGC~1345           &Sc   &  4.5  & 13.80 &       &      & 12.0 & 138\\
14 & NGC~1347           &Scd  &  5.3  & 13.55 &       &      &  0.6 & 105\\
15 & NGC~1353           &Sbc  &  3.0  & 11.73 &  8.11 & 3.62 &      &\\
16 & UGCA 077	        &Sdm  &  9.0  & 14.36 &       &      &  5.3 & 158\\
17 & ESO~482-~G~005 	&Sdm  &  8.0  & 14.33 &       &      &  5.4 & 184\\
18 & IC~1952            &Sbc  &  4.0  & 12.59 &  9.87 & 2.72 &  5.1 & 263\\
19 & ESO~548-~G~036	&S?   &       &       & 10.43 &      &  4.1 &\\ 
20 & IC~1953            &Sc   &  7.0  & 12.10 &  9.65 & 2.45 &  8.1 & 197\\
21 & NGC~1359           &Scm  &  9.0  & 12.37 & 11.17 & 1.20 & 25.7 & 223\\
22 & ESO~548-~G~043	&Sa   &       & 15.55 & 11.46 & 4.09 &      &\\
23 & ESO~548-~G~044	&S0/a & --1.3 & 14.11 & 10.37 & 3.74 &      &\\
24 & NGC~1371           &Sa   &  1.0  & 11.36 &  7.63 & 3.73 & 53.7 & 427\\
25 & NGC~1370           &E/S0 & --3.5 & 13.36 &  9.87 & 3.49 &      &\\ 
26 & ESO~548-~G~049	&S?   &       & 14.92 &       &      &      &\\
27 & IC~1962            &Sdm  &  8.0  & 13.74 &       &      &  5.5 & 171\\
28 & NGC~1377           &S0   & --2.0 & 13.29 &  9.72 & 3.57 &      &\\
29 & ESO~482-~G~013	&Sb   &       &       &       &      &  2.0 &  99\\
30 & NGC~1385           &Scd  &  6.0  & 11.14 &  8.57 & 2.57 & 22.6 & 204\\
31 & NGC~1383           &S0   & --2.0 & 13.25 &  9.44 & 3.81 &      &\\
32 & NGC~1390           &S0/a &  1.0  & 14.01 & 11.54 & 2.47 &  2.4 & 191\\
33 & NGC~1393           &S0   & --1.9 & 12.78 &  9.18 & 3.60 &      &\\
34 & NGC~1401           &S0   & --2.0 & 13.11 &  9.35 & 3.76 &      &\\
35 & ESO~548-~G~064	&S0   &       & 14.52 & 10.73 & 3.79 &      &\\
36 & ESO~548-~G~065	&Sa   &  0.7  & 14.56 & 12.90 & 1.66 &  2.0 & 145\\
37 & IC~0343            &S0   & --1.0 & 13.91 & 10.50 & 3.41 &      &\\
38 & NGC~1407           &E0   & --5.0 & 10.71 &  6.70 & 4.01 &      &\\
39 & ESO~482-~G~031	&dS0  &       &       & 12.37 &      &      &\\ 
40 & APMBGC~548-110-078 &dS0  &       &       &       &      &      &\\
41 & NGC~1415           &S0/a &  0.0  & 12.41 &  8.73 & 3.68 &      &\\
42 & NGC~1414           &Sbc  &  3.8  & 13.59 &       &      &  2.1 &\\
43 & ESO~548-~G~072	&S?   &       &       &       &      &      &\\
44 & NGC~1416           &E/S0 & --5.0 & 13.88 & 10.54 & 3.34 &      &\\
45 & IC~0345            &S0/a &       & 14.64 & 10.05 & 4.59 &      &\\
\end{tabular} 
\end{center} 
\end{table}

\newpage

\begin{table}
\begin{center}
\begin{tabular}{rllrrrrcr}
\hline
\hline
\bf{\ch \#} & \bf{\ch Name} & \bf{\ch Morph.} & \bf{\ch H.T.} & \bf{\ch B$_{0}^{T}$} & \bf{\ch K$^{T}$} & 
\bf{\ch B -- K} & \bf{\ch \HI Flux} & \bf{\ch W$_{20}$}\\ 
 & & & & (\ch mag) & (\ch mag) & (\ch mag) & (\ch Jy~km/s) & (\ch km/s) \\
\hline 
46 & ESO~482-~G~035	&Sab  &  2.1  & 13.42 &       &      &  4.2 & 177\\
47 & NGC~1422           &Sab  &  2.4  & 13.16 & 10.73 & 2.43 &      &\\
48 & IC~0346            &S0   & --0.8 & 13.37 &  9.78 & 3.59 &      &\\
49 & ESO~549-~G~002	&Im   & 10.0  & 14.53 &       &      &      \\
50 & MCG~-03-10-041     &Sdm  &  8.0  &       & 11.91 &      &  3.7 & 184\\ 
51 & NGC~1440           &S0   & --1.9 & 12.35 &  8.20 & 4.15 &      &\\
52 & NGC~1438           &S0/a &  0.0  & 12.94 &  9.62 & 3.32 &      &\\
53 & NGC~1452           &Sa   &  0.4  & 12.56 &  8.67 & 3.89 &      &\\ 
54 & ESO 549- G 018	&Sc   &  5.0  & 13.13 & 10.64 & 2.49 &      &\\
55 & NGC~1481           &S0   & --3.3 & 14.40 & 11.18 & 3.22 &      &\\
56 & NGC~1482           &S0/a & --0.8 & 13.01 &  8.48 & 4.53 &  5.5 & 131\\
57 & ESO~549-~G~035	&Sc   &       &       &       &      &  6.0 & 145\\
\hline \\
\multicolumn{9}{p{5.5in}}{\ch 1) The morphological types, Hubble types, and photographic B-band
magnitudes are from RC3 (Third reference catalog of galaxies; de Vaucouleurs et al. 1991)
provided by NED. The K-band magnitudes are from the {\it Two Micron All Sky Survey} (2MASS;
Jarrett et al. 2000). The single dish \HI flux integrals and \HI widths are from NED.}
\vspace{5in}
\end{tabular} 
\end{center} 
\end{table}

\newpage

\begin{table}
\begin{center}
\caption{Observational details} 
\vspace{0.05in}
\label{tab:Radobs}
\begin{tabular}{lccllr}
\hline
\hline
\bf{\#} & \bf{Date} & \bf{Pointing centre} & \bf{Galaxy}  & \bf{Sepa.} &\bf{rms}\\
  &     &$\alpha$ {\ch(J2000)} ~~~~~ $\delta$ \ch{(J2000)} & \bf{name}   & & (mJy/\\ 
  & (dd-mm-yy)          &({\ch hh~mm~ss})~~~~~ ($~\deg~~'~~~''$) & &(arc min) &beam)\\
\hline
1 &01-03-00 & 03 33 47.7~~ --19 29 31.0 &N~1359         & 0   & 1.8 \\
2 &27-10-00 & 03 35 01.4~~ --24 55 58.0 &N~1371         & 0   & 1.3 \\
3 &27-10-00 & 03 37 28.3~~ --24 30 05.0 &N~1385         & 0   & 1.3 \\
4 &15-12-00 & 03 22 06.5~~ --15 24 00.0 &N~1309         & 0   & 1.7 \\
5 &03-05-01 & 03 24 25.6~~ --21 32 39.0 &N~1325         & 0   & 1.9 \\
6 &06-05-01 & 03 23 25.1~~ --19 17 04.0 &S~0321.2-1929  & 0   & 1.3 \\
7 &17-05-01 & 03 54 33.4~~ --20 28 14.2 &N~1482         &2.3  & 1.0 \\ 
  &         & 			     &N~1481         &2.8  & 1.1 \\
  &         &                        &E~549-G035     &8.9  & 1.4 \\
8 &18-05-01 & 03 40 34.4~~ --18 30 23.0 &A~548-110-078 ($^{\times}$)  &4.6  & 1.0\\
  &         &                        &N~1407 ($^{\times}$)        &6.9  & 1.1\\
  &         &                        &I 0343 ($^{\times}$)        &7.5  & 1.1\\
  &         &                        &A 548-108-069 ($^{\times}$) &8.6  & 1.1\\
9 &19-05-01 & 03 27 02.3~~ --21 16 05.7 &N~1332 ($^{\times}$)        &4.1  & 1.1\\
  &         &                        &E~548-G016 ($^{\times}$)    &5.6  & 1.1\\
  &         &                        &N~1331 ($^{\times}$)        &5.9  & 1.1\\
10&20-05-01 & 03 45 02.9~~ --18 15 58.0 &N~1440 ($^{\times}$)         & 0   & 0.9\\
11&21-05-01 & 03 35 14.6~~ --20 22 25.0 &N~1370 ($^{\times}$)       & 0   & 0.9\\
12&22-05-01 & 03 36 39.1~~ --20 54 08.0 &N~1377 ($^{\times}$)        & 0   & 0.9\\
13&16-06-02 & 03 29 31.7~~ --17 46 43.0 &N~1345         & 0   & 1.5 \\
14&16-06-02 & 03 32 19.3~~ --17 43 07.0 &U~077          & 0   & 1.4 \\
15&17-06-02 & 03 35 37.3~~ --21 17 39.1 &I 1962         & 0   & 0.9 \\
  &         & 			     &E~548-G049     &5.1  & 1.1 \\
16&17-06-02 & 03 23 47.2~~ --19 45 15.0 &U 068          & 0   & 0.9 \\
17&18-06-02 & 03 41 13.9~~ --21 41 48.5 &N 1414         &4.0  & 0.9 \\
  &         &                        &N 1422         &4.1  & 1.0 \\
18&20-06-02 & 03 40 56.8~~ --22 38 24.5 &E 482-G031 ($^{\times}$)     &3.6  & 0.9\\
  &         &                        &N 1415         &4.5  & 0.9 \\
  &         &                        &N 1416  ($^{\times}$)       &4.9  & 0.9\\
19&22-06-02 & 03 29 21.1~~ --22 12 43.9 &E 548-G025 ($^{\times}$)    &6.2  & 1.2\\
  &         &                        &N 1347         &6.3  & 1.2 \\
20&23-06-02 & 03 33 26.5~~ --23 42 41.0 &I 1952         & 0   & 1.2 \\
21&23-06-02 & 03 33 34.5~~ --21 30 58.7 &I 1953         &2.8  & 1.4 \\
  &         &                        &E 548-G036     &3.3  & 1.4 \\
22&24-06-02 & 03 27 35.3~~ --21 13 42.0 &E 548-G021     & 0   & 0.9 \\
23&24-06-02 & 03 37 52.2~~ --19 00 30.0 &N 1390         & 0   & 1.1 \\
24&26-06-02 & 03 41 15.0~~ --23 50 10.0 &E 482-G035     & 0   & 1.2 \\
25&26-06.02 & 03 48 14.1~~ --21 28 28.0 &E 549-G018     & 0   & 1.4 \\
26&27-06-02 & 03 40 31.4~~ --19 24 40.0 &E 548-G065     &7.3  & 1.1 \\
  &         &                        &E 548-G072     &7.3  & 1.3 \\
  &         &                        &E 548-G064  ($^{\times}$)   &7.5  & 1.3\\
27&27-06-02 & 03 42 57.3~~ --19 01 12.0 &E 549-G002     & 0   & 1.0 \\
\end{tabular}
\end{center}
\end{table}

\begin{table}
\begin{center}
\begin{tabular}{lcclrr}
28&28-06-02 & 03 41 00.3~~ --19 27 19.0 &E 548-G077 ($^{\times}$)    & 0   & 1.2\\
29&28-06-02 & 03 43 35.5~~ --16 00 52.0 &M -03-10-041   & 0   & 1.2 \\

30&29-06-02 & 03 33 02.2~~ --24 07 58.0 &E 482-G005     & 0   & 1.1 \\
31&29-06-02 & 03 36 53.9~~ --24 54 46.0 &E 482-G013     & 0   & 1.1 \\
\hline
\multicolumn{6}{p{5.5in}}{\ch Notes:
(1) The bandwidth was 8~MHz except for pointings 5 and 6 where it 
was 4~MHz and for pointing 4 it  was 2~MHz. Therefore, the velocity resolutions were
$\sim13.4$~\km for all except for N~1325 and S~0321.2-1929 where it were $\sim6.7$~\km and for N~1309 it
was $\sim3.3$~\km. \newline
(2) The on-source integration time for observation was $\sim3$~hour except for
pointings 5,6,7,8,9  where it was $\sim7$ hour.\newline
(3) The centre frequency was 1412.6~MHz except for pointing 1 (1408.0~MHz), pointings 2 and 3
(1414.0~MHz), pointing 4 (1410.0~MHz), pointing 5 and 6 ($\sim1413$~MHz) and pointing 7 (1412.0~MHz).
\newline
(4) The column with the heading `Sepa' indicates the angular separation between the pointing centre and
the galaxy. \newline
(5) The $FWHM$ of the image cubes was $20''\times20''$ except for N~1309 and N~1325 where it was
$25''\times25''$ and for N~1371, E~482-G031, N~1415, and  N~1416 it was $30''\times30''$. \newline
(6) A cross against the galaxy name indicates an \HI non-detection.}

\end{tabular} 
\end{center} 
\end{table}

\newpage

\begin{table}
\begin{center}
\caption{Results from the GMRT \HI synthesis observations}
\label{tab:results}
\begin{tabular}{llllllll}
\hline
\hline
\ch{Galaxy Name}	& \ch{Type}	& \ch{V$_{sys}$}   & \ch{D$_{\HI}$}  & \ch{V$_{max}$} & \ch{M$_{\HI}$} & \ch{M$_{\HI}$/L$_{B}$} & \ch{$\sigma_{\HI}(\HI)$} \\
\ch{Opt. centre} &\ch{INCL.}&\ch{V$_{cent}$} &\ch{D$_{B25}$}  &\ch{V$_{flat}$} &\ch{M$_{K}$} &\ch{M$_{\HI}$/L$_{R}$}&\ch{$\sigma_{\HI}(opt)$} \\
\ch{Radio centre} &\ch{P.A.} &\ch{V$_{opt}$} &\ch{D$_{R25}$}  &\ch{$W_{50}/2$} &\ch{M$_{dyn}$} &\ch{M$_{\HI}$/L$_{K}$}&\ch{D$_{\HI}$/D$_{B25}$}\\
\ch{($\alpha(h,m,s),~\delta(\deg, ', '')$ J2000)}  & &\ch{(km/s)} &\ch{(kpc)}  &\ch{(km/s)} &\ch{(10$^{9}$M$_{\odot}$)} &\ch{(M$_{\odot}$/L$_{\odot}$)}
&\ch{(M$_{\odot}$/pc$^{2}$)}\\

\hline
ESO 482- G 005         &SBdm    &1923  & 32   &80.2  &1.08  &0.71  &1.34  \\
03 33 02.2 -24 07 58   &82\deg      &1918  &11.3  &80.2  & --   & --   &10.7  \\
03 33 2.15 -24 07 58   &264\deg     &1915  & --   &79.5  &8.423 & --   &2.83  \\
			& 	&	&	&	&	&	& \\
ESO 482- G 013         &Sb      &--    &12.9  &--    &0.578  &0.83 &4.44  \\
03 36 53.9 -24 54 46   &63      &1850  &7.59  &--    &0.604  &1.5  &12.8  \\
03 36 53.8 -24 54 46   &65      &1835  &7.17  &--    &1.11   &0.57 &1.7   \\
                        &       &       &       &       &       &       & \\
ESO 482- G 035         &SBab    &1883  &16.4  &118  &0.438  &0.13  &2.08  \\
03 41 15.0 -23 50 10   &49      &1884  &12.8  &118  &4.883  &0.2   &3.4   \\
03 41 14.7 -23 50 11   &185     &1890  &14.8  &115  &20.78  &0.054 &1.28  \\
                        &       &       &       &       &       &       & \\
ESO 548- G 021         &SBdm    &1690  & 20   &87.7  &0.425  &0.14 &1.36  \\
03 27 35.3 -21 13 42   &80      &1691  &13.4  &87.7  &0.507  & --  &  3   \\
03 27 35.2 -21 13 41.7 &64      &1668  & --   &83.3  &11.99  &0.5  &1.49  \\
                       &       &       &       &       &       &       &  \\
ESO 548- G 036         &S?      &--    &--   & --   &0.257  &0.13  & --   \\
03 33 27.6 -21 33 53   &--      &1507  &6.7  & --   &4.482  &--    &7.31  \\
03 33  28  -21 33 55.1 &--      &1480  &--   &60.5  &--     &0.034 &--    \\
                       &       &       &       &       &       &       &  \\
ESO 548- G 049         &S?      &1533  &14.9  & 71  &0.296  &0.34 &1.69   \\
03 35 28.1 -21 13 01   &71      &1533  &6.72  & 71  &--     &1.4  &8.36   \\
03 35 28.4 -21 13 7.01 &128     &1510  &6.27  &--   &3.93   &--   &2.22   \\
                    &       &       &       &       &       &       &     \\
ESO 548- G 065         &SBa     &1213  &14.4  &72.9  &0.289  &0.24 &1.77  \\
03 40 02.7 -19 22 00   &80      &1243  &10.1  &72.9  &--     &0.78 &3.63  \\
03 40 2.53 -19 21 56.8 &37      &1221  &--    &64.5  &6.214  &--   &1.43  \\
                       &       &       &       &       &       &       &  \\
ESO 548- G 072          &S?     &2045  &8.52  &44.3  &0.203  &0.77 &3.56  \\
03 41 00.3  -19 27 19   &74     &2052  &8.74  &44.3  &--     &--   &3.38  \\
03 41 0.795 -19 27  19  &51     &2034  &--    &48.0  &1.989  &--   &0.975 \\
                      &       &       &       &       &       &       &   \\
ESO 549- G 002         &IBm     &1110  & 9    &55.9  &0.189  &0.15 &2.97  \\
03 42 57.3 -19 01 12   &63      &1115  &8.74  &55.9  &--     &0.32 &3.15  \\
03 42 57.2 -19 01 11.1 &210     &1111  &--    &60.8  &3.167  &--   &1.03  \\
                      &       &       &       &       &       &       &   \\ 
ESO 549- G 018         &SABc    &1587  &15.5  &--  &0.429  &0.094 &2.26   \\
03 48 14.1 -21 28 28   &56      &1576  &17.5  &--  &12.37  &0.13  &1.79   \\
03 48  14  -21 28 28.9 &203     &1587  &17.5  &152 &--     &0.021 &0.89   \\

\end{tabular}
\end{center}
\end{table}

\begin{table}
\begin{center}
\begin{tabular}{llllllll}

ESO 549- G 035         &Sc      &1814  &14.7  &74.5  &0.526  &0.98 &3.11  \\
03 55 04.0 -20 23 01   &56      &1768  &9.41  &74.5  &--     &--   &7.57  \\
03 55 4.39 -20 23 0.92 &30      &1778  &--    &81.3  &6.057  &--   &1.56  \\
                      &       &       &       &       &       &       &   \\

IC 1952                &SBbc    &1820  &19.7  &134  &0.825  &0.11  &2.72  \\
03 33 26.7 -23 42 46   &71      &1823  &17.5  &134  &9.599  &0.23  &3.44  \\
03 33 26.4 -23 42 46.1 &319     &1812  &22.4  &122  &36.12  &0.052 &1.13  \\
                      &       &       &       &       &       &       &   \\
IC 1953                &SBd     &1863  &21.3  &150  &1.46  &0.12   &4.11  \\
03 33 41.9 -21 28 43   &37      &1863  &18.8  &150  &20.97 &0.2    &5.27  \\
03 33 42.2 -21 28 39.3 &129     &1867  &24.2  &142  &48.78 &0.042  &1.13  \\
                      &       &       &       &       &       &       &   \\
IC 1962                &SBdm    &1806  &26.6  &82.3  &0.855  &0.33 &1.54  \\
03 35 37.4 -21 17 33   &80      &1811  &18.1  &82.3  &0.7401 &1.2  &3.31  \\
03 35 37.8 -21 17  38  &358     &1806  &18.4  &69.1  &14.26  &0.69 &1.47  \\
                      &       &       &       &       &       &       &   \\
MCG -03-10-041         &SBdm    &1207  &19.2  &110  &0.938  &--   &3.25   \\
03 43 35.5 -16 00 52   &57      &1217  &13.4  &110  &2.112  &0.77 &6.61   \\
03 43 35.4 -16 0 51.3  &343     &1215  &13.9  &109  &18.9   &0.27  &1.43  \\
                      &       &       &       &       &       &       &   \\
NGC 1309               &SAbc    &2134  &24.1  &164  &3.33  &0.22 &7.28    \\
03 22 06.5 -15 24 00   &20      &2134  &14.8  &164  &16.44 &0.53 &19.4    \\
03 22 5.89 -15 23 59.9 &210     &2135  &13.4  &168  &46.35 &0.12 &1.63    \\
                      &       &       &       &       &       &       &   \\
NGC 1325               &SAbc    &1593  & 44   &158  &3.14  &0.15 &2.06    \\
03 24 25.6 -21 32 39   &71      &1594  &31.6  &158  &27.49 &0.27 &4.01    \\
03 24 25.6 -21 32 35.9 &232     &1589  &38.5  &169  &91.69 &0.069 &1.39   \\
                      &       &       &       &       &       &       &   \\
NGC 1345               &SBc     &1530  &23.9  &116  &1.57  &0.64  &3.51   \\
03 29 31.7 -17 46 43   &34      &1536  &10.1  &116  &2.766 &1.3   &19.7   \\
03 29 31.8 -17 46 40.3 &88      &1529  &8.51  &112  &15.87 &0.34  &2.37   \\
                     &       &       &       &       &       &       &    \\
NGC 1347               &SBcd    &1758  &12.9  &96.9  &0.438  &0.14  &3.35 \\
03 29 41.8 -22 16 45   &26      &1784  &7.06  &96.9  &3.364  &0.38  &11.2 \\
03 29 41.7 -22 16 44.7 &328     &1759  &8.51  &--   &7.685  &0.078 &1.83 \\
                     &       &       &       &       &       &       &    \\
NGC 1359               &SBcm    &1973  &--   & --  &3.44  &0.37  & --     \\
03 33 47.7 -19 29 31   &53      &1980  &--   & --  &5.816 & 1    & --     \\
03 33 47.2 -19 29 19.1 &325     &1966  &17.9 &108  & --   &0.35  & --     \\
                     &       &       &       &       &       &       &    \\
NGC 1371               &SABa    &1467  & 61  &260   &4.2   &0.18  &1.44   \\
03 35 01.4 -24 55 58   &49      &1469  &37.6  &260  &61.55 &0.24  &3.78   \\
03 35 1.89 -24 55 58.6 &136     &1471  &34.5  &246  &295.1 &0.041 &1.62   \\

\end{tabular}
\end{center}
\end{table}

\begin{table}
\begin{center}
\begin{tabular}{llllllll}

NGC 1385               &SBcd    &1493  &19.8  &138  &2.67  &0.093 &8.69   \\
03 37 28.3 -24 30 05   &40      &1503  &22.8  &138  &33.66 &0.18  &6.51   \\
03 37 27.8 -24 30 6.68 &181     &1493  &30.5  &140  &50.77 &0.048 &0.866  \\
                     &       &       &       &       &       &       &    \\
NGC 1390               &SB0/a    &1207  &15.2  &118   &0.312  &0.15 &1.72 \\
03 37 52.2 -19 00 30   &60       &1221  &9.41  &118   &1.845  &0.38 &4.49 \\
03 37 52.1 -19 0 29.6  &24       &1207  &8.96  &97.2  &15.27  &0.1  &1.62 \\
                     &       &       &       &       &       &       &    \\
NGC 1414               &SBbc    &1695  &19.9  &78.1  &0.577  &0.19 &1.85  \\
03 40 57.0 -21 42 47   &80      &1660  &11.6  &78.1  & 1.4   &1.6  &5.44  \\
03 40 57.3 -21 42 49.3 &357     &1681  &11.2  &95.4  &8.226 &0.25 &1.71   \\
                    &       &       &       &       &       &       &    \\
NGC 1415               &SAB0/a  &1585  &--    &158  &1.11  &0.13 & --     \\
03 40 56.8 -22 33 52   &--      &1576  &--    &158  &36.34 &0.089 &--     \\
03 40 57   -22 33 49.7 &--      &1585  &32.7  &167  &--    &0.018 & --    \\ 
                    &       &       &       &       &       &       &     \\ 
NGC 1422               &SBab    &1630  &15.4  &73.8  &0.312  &0.07 &1.68  \\
03 41 31.2 -21 40 53   &80      &1660  &14.8  &73.8  &3.722  &0.34 &1.82  \\
03 41 31.2 -21 40 54.4 &65      &1637  &18.4  &54.2  &9.34   &0.05 &1.04  \\
                    &       &       &       &       &       &       &     \\
NGC 1481               &S0      &--    &3.4   &--   &0.079  &0.06  &8.8 \\
03 54 29.0 -20 25 38   &--      &--    &6.7   &--   &2.2    &--    &2.3 \\
		       &--      &1733  &--    &--   &--     &0.02  &0.51 \\
		    & 	    &     &	     &	    &	  &	& \\
NGC 1482	       &S0/a    &--    &9.0   &--   &0.428  &0.07  &5.0 \\
03 54 38.9 -20 30 08   &--      &--    &16.8  &--   &28.0   &--    &1.45 \\
 		       &--      &1916  &--    &--   &--     &0.01  &0.54 \\
		      & 	 &	&	&	&	&	& \\
UGCA 068               &SABcdm  &1852  &16.9  &87.6  &0.638  &0.21 &2.84  \\
03 23 47.2 -19 45 15   &34      &1843  &11.4  &87.6  &0.931  &0.97 &6.22  \\
03 23 47.1 -19 45 15.2 &35      &1838  &10.3  &84.1  &10.17  &0.41 &1.48  \\
                    &       &       &       &       &       &       &     \\
UGCA 077               &SBdm    &1963  &22.1  &77.9  &1.28  &0.87 &3.34   \\
03 32 19.3 -17 43 07   &66      &1956  &11.4  &77.9  &--    &2.5  &12.4   \\
03 32 19.1 -17 43 10.2 &149     &1961  &12.1  &74.5  &8.042 &--   &1.93   \\
                    &       &       &       &       &       &       &     \\
SGC 0321.2-1929        &IBm     &1552  &13.1  &56.4  &0.341  &--   &2.53  \\
03 23 25.1 -19 17 04   &24      &1556  &--    &56.4  &--     &5.3  &--    \\
03 23 25.2 -19 17 5.33 &175     &1545  &--    &54.7  &--     &--   &--    \\

\hline
\end{tabular}
\end{center}
\end{table}

\newpage

\begin{figure} \centering \includegraphics[width=12cm]{ESO_482-_G_005.rad-atlas2.epsi} \end{figure} \newpage
\begin{figure} \centering \includegraphics[width=12cm]{ESO_482-_G_013.rad-atlas2.epsi} \end{figure} \newpage
\begin{figure} \centering \includegraphics[width=12cm]{ESO_482-_G_035.rad-atlas2.epsi} \end{figure} \newpage
\begin{figure} \centering \includegraphics[width=12cm]{ESO_548-_G_021.rad-atlas2.epsi} \end{figure} \newpage
\begin{figure} \centering \includegraphics[width=12cm]{ESO_548-_G_036.rad-atlas2.epsi} \end{figure} \newpage
\begin{figure} \centering \includegraphics[width=12cm]{ESO_548-_G_049.rad-atlas2.epsi} \end{figure} \newpage
\begin{figure} \centering \includegraphics[width=12cm]{ESO_548-_G_065.rad-atlas2.epsi} \end{figure} \newpage
\begin{figure} \centering \includegraphics[width=12cm]{ESO_548-_G_072.rad-atlas2.epsi} \end{figure} \newpage
\begin{figure} \centering \includegraphics[width=12cm]{ESO_549-_G_002.rad-atlas2.epsi} \end{figure} \newpage
\begin{figure} \centering \includegraphics[width=12cm]{ESO_549-_G_018.rad-atlas2.epsi} \end{figure} \newpage
\begin{figure} \centering \includegraphics[width=12cm]{ESO_549-_G_035.rad-atlas2.epsi} \end{figure} \newpage
\begin{figure} \centering \includegraphics[width=12cm]{IC_1952.rad-atlas2.epsi} \end{figure} \newpage
\begin{figure} \centering \includegraphics[width=12cm]{IC_1953.rad-atlas2.epsi} \end{figure} \newpage
\begin{figure} \centering \includegraphics[width=12cm]{IC_1962.rad-atlas2.epsi} \end{figure} \newpage
\begin{figure} \centering \includegraphics[width=12cm]{MCG_-03-10-041.rad-atlas2.epsi} \end{figure} \newpage
\begin{figure} \centering \includegraphics[width=12cm]{NGC_1309.rad-atlas2.epsi} \end{figure} \newpage
\begin{figure} \centering \includegraphics[width=12cm]{NGC_1325.rad-atlas2.epsi} \end{figure} \newpage
\begin{figure} \centering \includegraphics[width=12cm]{NGC_1345.rad-atlas2.epsi} \end{figure} \newpage

\begin{figure} \centering \includegraphics[width=12cm]{NGC_1347.rad-atlas2.epsi} \end{figure} \newpage
\begin{figure} \centering \includegraphics[width=12cm]{NGC_1359.rad-atlas2.epsi} \end{figure} \newpage
\begin{figure} \centering \includegraphics[width=12cm]{NGC_1371.rad-atlas2.epsi} \end{figure} \newpage
\begin{figure} \centering \includegraphics[width=12cm]{NGC_1385.rad-atlas2.epsi} \end{figure} \newpage
\begin{figure} \centering \includegraphics[width=12cm]{NGC_1390.rad-atlas2.epsi} \end{figure} \newpage
\begin{figure} \centering \includegraphics[width=12cm]{NGC_1414.rad-atlas2.epsi} \end{figure} \newpage
\begin{figure} \centering \includegraphics[width=12cm]{NGC_1415.rad-atlas.epsi} \end{figure} \newpage
\begin{figure} \centering \includegraphics[width=12cm]{NGC_1422.rad-atlas2.epsi} \end{figure} \newpage
\begin{figure} \centering \includegraphics[width=12cm]{NGC_1481.rad-atlas.epsi} \end{figure} \newpage
\begin{figure} \centering \includegraphics[width=12cm]{NGC_1482.rad-atlas.epsi} \end{figure} \newpage
\begin{figure} \centering \includegraphics[width=12cm]{SGC_0321.2-1929.rad-atlas2.epsi} \end{figure} \newpage
\begin{figure} \centering \includegraphics[width=12cm]{UGCA_068.rad-atlas2.epsi} \end{figure} \newpage
\begin{figure} \centering \includegraphics[width=12cm]{UGCA_077.rad-atlas2.epsi} \end{figure} \newpage

\end{document}